\documentclass[10pt, twocolumn]{IEEEtran}

\usepackage{amsmath}
\usepackage{amsfonts}
\usepackage{amssymb,bm}
\usepackage{amsthm}
\usepackage{algorithm}
\usepackage{algpseudocode}
\usepackage{cite}
\usepackage{graphicx}
\usepackage{epstopdf}
\usepackage[caption=false,font=footnotesize]{subfig}
\usepackage{fixltx2e}
\usepackage{color}
\usepackage{balance}

\allowdisplaybreaks

\begin{document}
%
\title{Sum-Rate Maximization for Linearly Precoded Downlink Multiuser MISO Systems with Partial CSIT: A Rate-Splitting Approach}
\author{
Hamdi~Joudeh and Bruno~Clerckx
\thanks{This work is partially supported by the U.K. Engineering and Physical Sciences Research Council (EPSRC) under grant EP/N015312/1.
A preliminary version of this paper was presented at the IEEE International Conference on Communications (ICC), London, U.K., June 2015.}
\thanks{H.~Joudeh is with the Communications and Signal Processing group, Department of Electrical and Electronic Engineering, Imperial College London, London SW7 2AZ, U.K. (email: hamdi.joudeh10@imperial.ac.uk)}
\thanks{B.~Clerckx is with the Communications and Signal Processing group, Department of Electrical and Electronic Engineering, Imperial College London, London SW7 2AZ, U.K., and also with the School of Electrical Engineering, Korea University, Seoul 136-701,
Korea (e-mail: b.clerckx@imperial.ac.uk).}
}
\maketitle

\begin{abstract}
This paper considers the Sum-Rate (SR) maximization problem in downlink MU-MISO systems under imperfect Channel State Information at the Transmitter (CSIT).
Contrary to existing works, we consider a rather unorthodox transmission scheme.
In particular, the message intended to one of the users is split into two parts: a common part which can be recovered by all users, and a private part recovered by the corresponding user.
On the other hand, the rest of users receive their information through private messages.
This Rate-Splitting (RS) approach was shown to boost the achievable Degrees of Freedom (DoF) when CSIT errors decay with increased SNR.
In this work, the RS strategy is married with linear precoder design and optimization techniques to achieve a maximized Ergodic SR (ESR) performance over the entire range of SNRs.
Precoders are designed based on partial CSIT knowledge by solving a stochastic rate optimization problem using means of Sample Average Approximation (SAA) coupled with the Weighted Minimum Mean Square Error (WMMSE) approach.
Numerical results show that in addition to the ESR gains, the benefits of RS also include relaxed CSIT quality requirements and enhanced achievable rate regions compared to conventional transmission with NoRS.
\end{abstract}


\IEEEpeerreviewmaketitle

\section{Introduction}
\newcounter{Theorem_Counter}
\newcounter{Proposition_Counter} 
\newcounter{Lemma_Counter} 
\newcounter{Remark_Counter} 
\newcounter{Assumption_Counter}
\newcounter{Definition_Counter}
\IEEEPARstart{T}{he} utilization of multiple antennas at the Base Station (BS) combined with multiple single-antenna mobile devices tremendously increases the spectral efficiencies of wireless networks. High multiplexing gains are realized with far less restrictions on the scattering environment compared to point-to-point Multiple Input Multiple Output (MIMO) systems \cite{Clerckx2013}.
However, this comes with a price of higher restrictions imposed on the quality of the Channel State Information (CSI) required at the BS, specifically in the Downlink (DL) mode.
This stems from the necessity to deal with the interference through preprocessing at the BS, as receivers cannot coordinate.
While precise CSI at the Receivers (CSIR) can be obtained through DL training, the ability to provide highly accurate and up-to-date CSI at the Transmitter (CSIT) remains questionable.
Therefore, considerable effort has been devoted to the characterization and improvement of the performance in the presence of CSIT uncertainties.

It is well established that linear transmission strategies, e.g. Zero Forcing Beamforming (ZF-BF), achieve the optimum Degrees of Freedom (DoF) of
the MISO Broadcast Channel (BC) under perfect CSIT \cite{Caire2003,Jindal2005}.
It is also possible to maintain the full DoF in the presence of imperfect CSIT given that the errors decay with increased Signal to Noise Ratio (SNR) as $O(\mathrm{SNR}^{-1})$ \cite{Caire2007,Jindal2006,Caire2010}.
For example, it has been shown that employing ZF-BF in limited channel feedback systems achieves the maximum DoF as
long as the feedback rate increases linearly with the SNR in dB, satisfying an error decay rate of $O(\mathrm{SNR}^{-1})$ \cite{Jindal2006}.
However, maintaining such quality can be exhausting in terms of resources required for CSIT acquisition.
Furthermore, the CSIT accuracy is expected to drop due to the Doppler effect, feedback delays and/or mismatches across multiple uses of the channel.
While understanding the MISO-BC capacity region under CSIT imperfections is still far from complete,
considerable progress has been made towards characterizing the achievable and optimum DoF -- as an asymptotic tractable alternative -- for a variety of imperfect CSIT scenarios \cite{Yang2013,Hao2013,Tandon2013,Chen2013}.
In this work, we focus on the case where imperfect instantaneous CSIT is available, with errors decaying as $O(\mathrm{SNR}^{-\alpha})$ for some constant $\alpha \in [0,1]$.
Such errors are detrimental to the DoF achieved using conventional transmission schemes initially designed assuming perfect CSIT. For example,
ZF-BF achieves a fraction $\alpha$ of the full DoF \cite{Jindal2006,Caire2010}.
Therefore we consider a rather unconventional Rate-Splitting (RS) transmission strategy.
The message intended to one of the users (labeled as the RS-user) is split into two parts: a common part decoded by all users, and a private part decoded by the corresponding user.
On the other hand, the information intended to the remaining users is sent through private messages.
At the receivers, the common message is decoded first by treating all private signals as noise.
Then, each receiver decodes its private message after removing the common message from the received signal.
This RS strategy, inspired by similar techniques used for the Interference Channel (IC) \cite{ElGamal2011}, was recently shown to achieve a DoF gain of $1-\alpha$ over conventional transmission (relying solely on the transmission of private messages and termed NoRS here) in the MISO-BC \cite{Yang2013,Hao2013}.
An interpretation of this result in the context of limited feedback and Random Vector Quantization (RVQ) of the channel vectors is given in \cite{Hao2015}, where the RS rate performance is analysed and compared to its NoRS counterpart.
While RS in the MISO-BC is fairly understood from an information theoretic point of view (at least in the DoF sense),
it is considerably less treated and analysed in the transceiver/precoder design and optimization literature.
For example, none of the aforementioned works considers an optimized design of precoders.
One reason is that theoretical analysis is made possible by simpler designs, e.g. random precoding for the common message and ZF-BF for private messages \cite{Hao2015}, as it is easier to derive closed-form expressions of performance metrics.
Such closed-forms are usually sufficient for DoF analysis, where the asymptotically high SNR regime is considered.
However, this is not the case for finite SNRs where the design of precoders is highly influential.
Precoder optimization for Multi-User (MU) MISO/MIMO systems has been considered in a number of works under a variety of metrics, design objectives and constraints, and  CSI assumptions  \cite{Stojnic2006,Christensen2008,Shi2008,Shenouda2008,Vucic2009a,Bogale2011,Bashar2014,Negro2012,Fritzsche2013,Wu2014}.
In general, problems that involve Sum-Rate (SR) expressions in such setups are known to be very challenging to solve in their raw forms due to the non-convexity arising from decentralized receivers with coupled rate expressions\footnote{This can be resolved by resorting to asymptotic regimes, e.g. high SNR or large antenna arrays \cite{Bjoernson2014}. From a design and optimization point of view, this work is mainly concerned with finite SNRs and small antenna arrays.}.
Nevertheless, the novel relationship proposed in \cite{Christensen2008} enables the reformulation of SR problems into equivalent special forms of Weighted Mean Square Error (WMSE) problems. This unveils a block-wise convexity property which can be exploited using Alternating Optimization (AO).

\emph{Contributions}:
With the aim of maximizing the SR performance in MU-MISO systems with partial
CSIT, this paper marries the RS strategy with precoder design and optimization
techniques.
In doing so, we generalize previous works on SR maximization under imperfect CSIT in several directions:
\begin{itemize}
\item Firs, we consider a general channel fading model where the channel state, and its corresponding estimate at the BS, are allowed to change throughout the transmission according to some stationary process.
    This model contains as special cases the Rician-fading scenario in \cite{Bashar2014,Negro2012,Fritzsche2013} where a fixed channel estimate (e.g. line-of-sight) throughout the transmission was considered, and the fast-fading scenario in \cite{Razaviyayn2013a} where the instantaneous channel estimate is completely suppressed.
    A feasible solution is defined as a sequence of precoders, with one for each incoming channel estimate.
\item Second, the adoption of the RS strategy ensures that the proposed design performs at least as well as the corresponding NoRS-based design  across the entire range of SNRs. This follows from the fact that NoRS can be seen as a special (or restricted) case of RS.
    We demonstrate that RS designs outperform conventional NoRS designs at high SNRs through DoF analysis.
\end{itemize}

In order to average out the effects of CSIT errors, we consider the Ergodic SR (ESR) achieved over a long
sequence of fading states as the main performance metric \cite{Caire2007a}.
Given partial CSIT, the ESR is maximized by updating the precoding matrix according to the incoming channel state estimate such that a conditional Average SR (ASR) metric is maximized.
This metric corresponds to the average performance w.r.t CSIT errors for a given channel estimate, and can be computed by the BS using imperfect instantaneous knowledge.
To solve the stochastic ASR problem, it is first converted into a deterministic counterpart using the Sampling Average Approximation (SAA) method \cite{Shapiro2009}.
The deterministic problem is then solved using the Weighted Minimum MSE (WMMSE) approach \cite{Christensen2008}.
A simplified method to approximate the stochastic problem based on the conservative method in \cite{Negro2012,Bashar2014} is also proposed, and it limitations are discussed.
The benefits of employing the RS strategy are demonstrated through simulations. These include enhanced rate performances and relaxed CSIT quality requirements.
Moreover, the two-user RS rate region is numerically obtained by solving a sequence of Weighted ASR (WASR) problems, where the splitting procedure is extended to both users.
Examining such achievable rate region brings us one step closer to understanding the behaviour of the MISO-BC's capacity region under partial CSIT.
It should be noted that part of the results presented here were reported in a preliminary version of this paper \cite{Joudeh2015}.
Since then, a number of RS design problems have been addressed, namely: achieving max-min fairness \cite{Joudeh2015a}, robust transmission under bounded CSIT errors \cite{Joudeh2016a}, and the application to massive MIMO \cite{Dai2016}.
In a recent work, a topological RS strategy is proposed for MISO networked systems \cite{Hao2016}.

\emph{Organization}:
The rest of the paper is organized as follows. The system model and CSIT assumptions are described in Section \ref{Section_System_Model}.
In Section \ref{Section_Problem_Statement}, the problem is formulated and DoF analysis is carried out.
The SAA based WMMSE algorithm is proposed in Section \ref{Section_ASR_AWSMSE}, then the conservative WMMSE algorithm is describe in  Section \ref{Section_Conservative}.
Simulation results are presented in Section \ref{Section_Numerical_Results}, and Section \ref{Section_conclusion} concludes
the paper.

\emph{Notation}: Boldface uppercase letters denote matrices, boldface lowercase letters denote column vectors and standard letters denote scalars. The superscrips $(\cdot)^{T}$ and $(\cdot)^{H}$ denote transpose and conjugate-transpose (Hermitian) operators, respectively. $\mathrm{tr}(\cdot)$ and $\mathrm{diag(\cdot)}$ are the trace and diagonal entries respectively. $\|\cdot\|$ is the Euclidian norm. $\mathrm{E}_{\mathrm{X}}\{\cdot\}$ denotes the expectation w.r.t the random variable $\mathrm{X}$.
%
\section{System Model}
\label{Section_System_Model}
Consider a MU-MISO system operating in DL, where a BS equipped with $N_{\mathrm{t}}$ antennas serves a set of single-antenna users $\mathcal{K} \triangleq  \{1,\ldots,K\}$, where $K \leq N_{\mathrm{t}}$.
The signal received by the $k$th user in a given channel use (time or frequency) is described as
\begin{equation}\label{Eq_yk}
    y_{k}=\mathbf{h}_{k}^{H}\mathbf{x}+n_{k}
\end{equation}
where
$\mathbf{h}_{k} \in \mathbb{C}^{N_{\mathrm{t}}}$ is the channel vector between the BS and the $k$th user,
$\mathbf{x}\in\mathbb{C}^{N_{\mathrm{t}}}$ is the transmit signal, and $n_{k} \thicksim \mathcal{CN} ( 0 , \sigma^{2}_{\mathrm{n},{k}} )$ is the Additive White Gaussian Noise (AWGN).
The input signal is subject to the power constraint $\mathrm{E}\{\mathbf{x}^{H}\mathbf{x}\} \leq P_{\mathrm{t}}$.
Without loss of generality, we assume equal noise variances across users, i.e. $\sigma_{\mathrm{n},{k}}^{2}=\sigma_{\mathrm{n}}^{2}, \ \forall k \in \mathcal{K}$. The transmit SNR writes as $\mathrm{SNR} \triangleq P_{\mathrm{t}} / \sigma_{\mathrm{n}}^{2}$.
Moreover, $\sigma_{\mathrm{n}}^{2}$ is non-zero and fixed. Hence, $\mathrm{SNR} \rightarrow \infty$ is equivalent to $P_{\mathrm{t}}  \rightarrow \infty$.
\subsection{Channel State Information}
\label{Subsection_CSIT}
We assume a fading model where the channel state, given by $\mathbf{H} \triangleq [\mathbf{h}_{1},\ldots,\mathbf{h}_{K}]$, varies during the transmission according to an ergodic stationary process with probability density
$f_{\mathrm{H}}\big( \mathbf{H} \big)$.
Receivers are assumed to estimate and track their channel vectors with high accuracy, i.e. perfect CSIR.
The BS on the other hand has an imperfect instantaneous channel estimate given by $\widehat{\mathbf{H}}\triangleq[\widehat{\mathbf{h}}_{1},\ldots,\widehat{\mathbf{h}}_{K}]$. This is typically obtained through Uplink (UL) training in Time Division Duplex (TDD) systems \cite{Hassibi2003} or quantized feedback in Frequency Division Duplex (FDD) systems \cite{Love2008}.
Hence, the joint fading process is characterized by the joint distribution of
$\big\{\mathbf{H},\widehat{\mathbf{H}}\big\}$, assumed to be stationary and ergodic \cite{Caire2007a}.
For a given estimate, the estimation error matrix is denoted by $\widetilde{\mathbf{H}}\triangleq [\widetilde{\mathbf{h}}_{1},\ldots,\widetilde{\mathbf{h}}_{K}]$,
from which we write the relationship $\mathbf{H} = \widehat{\mathbf{H}} + \widetilde{\mathbf{H}}$.
The CSIT error is characterized by the conditional density
$f_{\mathrm{H}\mid\widehat{\mathrm{H}}}\big( \mathbf{H} \mid \widehat{\mathbf{H}} \big)$.
Taking each user separately, the marginal density of the $k$th channel conditioned on its estimate writes as  $f_{\mathrm{h}_{k}\mid\widehat{\mathrm{h}}_{k}}\big( \mathbf{h}_{k} \mid  \widehat{\mathbf{h}}_{k}  \big)$.
The mean of the distribution is assumed to be given by the estimate, i.e.
$\mathrm{E}_{\mathrm{h}_{k} \mid\widehat{\mathrm{h}}_{k}} \{\mathbf{h}_{k} \mid \widehat{\mathbf{h}}_{k} \} = \widehat{\mathbf{h}}_{k}$, and
$\mathrm{E}_{\mathrm{h}_{k}\mid\widehat{\mathrm{h}}_{k}} \{\mathbf{h}_{k}\mathbf{h}_{k}^{H} \mid \widehat{\mathbf{h}}_{k} \} = \widehat{\mathbf{h}}_{k}\widehat{\mathbf{h}}_{k}^{H} + \mathbf{R}_{\mathrm{e},k}$, where $ \mathbf{R}_{\mathrm{e},k}$ is the $k$th user's CSIT error covariance matrix assumed to be independent of $\widehat{\mathbf{h}}_{k}$.
The CSIT quality is allowed to scale with SNR \cite{Jindal2006,Yang2013}.
In particular, the maximum entry of $\mathrm{diag}\left( \mathbf{R}_{\mathrm{e},k} \right)$ scales as $O(P_{\mathrm{t}}^{-\alpha})$, where $\alpha \in [0,\infty)$ is some constant.
Equivalently, we write $\sigma_{\mathrm{e},k}^{2} = O(P_{\mathrm{t}}^{-\alpha})$, where
$\sigma_{\mathrm{e},k}^{2} \triangleq
\mathrm{E}_{\widetilde{\mathrm{h}}_{k}} \big\{ \| \widetilde{\mathbf{h}}_{k} \|^{2} \big\} = \mathrm{tr}(\mathbf{R}_{\mathrm{e},k})$ is the corresponding average CSIT error power.
The constant $\alpha \triangleq \lim_{P_{\mathrm{t}}\rightarrow\infty} -\frac{\log(\sigma_{\mathrm{e},k}^{2})}{\log(P_{\mathrm{t}})}$
is known as the quality scaling factor (or exponent), which quantifies the CSIT quality as SNR grows large.
For example, $\alpha \rightarrow \infty$ corresponds to perfect CSIT as
$\sigma_{\mathrm{e},1}^{2},\ldots,\sigma_{\mathrm{e},K}^{2} \rightarrow  0$.
The opposite extreme of $\alpha =0$ represents a fixed quality w.r.t SNR, e.g. a constant number of quantization (and hence feedback) bits in FDD systems.
A finite non-zero $\alpha$ corresponds to CSIT quality that improves with increased SNR, e.g. by increasing the number of feedback bits.
The exponent is truncated such that $\alpha \in [0,1]$, which is customary in DoF analysis as $\alpha = 1$ corresponds to perfect CSIT in the DoF sense\cite{Yang2013}.
It is worth highlighting that $\alpha$ assumes various practical interpretations, e.g. in addition to the aforementioned channel quantization and feedback interpretation \cite{Jindal2006,Hao2015}, it can also correspond to the Doppler process in delayed/outdated CSIT \cite{Caire2010,Yang2013}.
\subsection{Rate-Splitting and Transmit Signal Model}
\label{Subsection_Rate_Splitting}
The BS wishes to communicate the independent messages $W_{\mathrm{t},1},\ldots,W_{\mathrm{t},K}$, drawn from the message sets $\mathcal{W}_{\mathrm{t},1},\ldots,\mathcal{W}_{\mathrm{t},K}$, to users $1,\ldots,K$ respectively.
In conventional linearly precoded systems, $W_{\mathrm{t},1},\ldots,W_{\mathrm{t},K}$ are independently encoded into the private symbol streams $s_{1},\ldots,s_{K}$ respectively.
Symbols are mapped to the transmit antennas through a precoding matrix denoted by
$\mathbf{P}_{\mathrm{p}} \triangleq  \big[\mathbf{p}_{1},\ldots,\mathbf{p}_{K}\big]$, where $\mathbf{p}_{k}\in\mathbb{C}^{N_{\mathrm{t}}}$ is the $k$th precoding vector.
This yields the linear transmit signal model described as $\mathbf{x} = \sum_{k=1}^{K}\mathbf{p}_{k}s_{k}$.
At each receiver, the desired stream is decoded while treating interference from all other streams as noise.
It is, therefore, the BS's duty to minimize MU interference through a proper design of $\mathbf{P}_{\mathrm{p}}$.
Unfortunately, this is not possible when the BS experiences high CSIT uncertainty.
Hence, we resort to the unconventional RS transmission model, where part of the interference is broadcasted such that it is decoded and cancelled by all users before decoding their own streams \cite{Hao2015}.
The RS scheme is described as follows. The message intended to the $k_{\mathrm{RS}}$th user (the RS-user) is split into a common part
$W_{\mathrm{c}} \in \mathcal{W}_{\mathrm{c}}$ and a private part $W_{k_{\mathrm{RS}}} \in \mathcal{W}_{k_{\mathrm{RS}}}$, where
$\mathcal{W}_{\mathrm{c}} \times \mathcal{W}_{k_{\mathrm{RS}}} = \mathcal{W}_{\mathrm{t},k_{\mathrm{RS}}}$.
$W_{\mathrm{c}}$ is encoded into the stream $s_{\mathrm{c}}$ using a common (or public) codebook, and hence is decoded by all users, while $W_{\mathrm{t},1},\ldots,W_{k_{\mathrm{RS}}},\ldots,W_{\mathrm{t},K}$ are encoded into the private stream $s_{1},\ldots,s_{K}$ in the conventional manner.
The $K+1$ streams are linearly precoded using $\mathbf{P} \triangleq  \big[\mathbf{p}_{\mathrm{c}},\mathbf{p}_{1},\ldots,\mathbf{p}_{K}\big]$, where $\mathbf{p}_{\mathrm{c}}\in\mathbb{C}^{N_{\mathrm{t}}}$ is the common precoder.
The resulting transmit signal writes as
\begin{equation}\label{Eq_x}
  \mathbf{x} = \mathbf{P}\mathbf{s} = \mathbf{p}_{\mathrm{c}}s_{\mathrm{c}} + \sum_{i=1}^{K}\mathbf{p}_{i}s_{i}
\end{equation}
%
where $\mathbf{s} \triangleq  [s_{\mathrm{c}},s_{1},\ldots,s_{K}]^{T} \in\mathbb{C}^{K+1}$ groups the symbols in a given channel use.
Assuming that $\mathrm{E}\{\mathbf{s}\mathbf{s}^{H}\}=\mathbf{I}$, the transmit power constraint reduces to
$\mathrm{tr}\big(\mathbf{P}\mathbf{P}^{H}\big) \leq P_{\mathrm{t}}$.
When $|W_{\mathrm{c}}| = 0$, i.e. no splitting is carried out, it is natural to allocate zero power to the common precoder, and the signal in \eqref{Eq_x} simply reduces to the conventional transmit signal.
While the model in \eqref{Eq_x} can be seen as a super-position of MU beamforming \cite{Schubert2004} and multicast beamforming \cite{Sidiropoulos2006}\footnote{A BC with both private messages and a common message from an information theoretic point of view \cite{ElGamal2011}.},
the transmission of a common message has a fundamentally different purpose here.
In particular, multicast transmission sends information requested by, and intended to, all users in the system.
On the other hand, the common message in RS encapsulates part of the RS-user's private message, which is decoded by all users in the system for interference mitigation and performance enhancement purposes as we see later in this paper\footnote{Note that decoding other user(s) data at the physical layer does not necessarily breach secrecy as encryption may be implemented at higher layers.
}.
%
\subsection{SINRs and Rates}
\label{Subsection_SINR_Rate}
At the $k$th receiver, the average receive power for a given channel state is written as
\begin{equation}
\label{Eq_T_c_k}
T_{\mathrm{c},k} \triangleq \mathrm{E} \big\{|y_{k}|^{2} \big\} = \overbrace{|\mathbf{h}_{k}^{H}\mathbf{p}_{\mathrm{c}}|^{2}}^{S_{\mathrm{c},k}} + \underbrace{ \overbrace{|\mathbf{h}_{k}^{H}\mathbf{p}_{k}|^{2}}^{S_{k}} + \overbrace{\sum_{i\neq k} |\mathbf{h}_{k}^{H}\mathbf{p}_{i}|^{2} + \sigma_{\mathrm{n}}^{2}}^{I_{k}}}_{I_{\mathrm{c},k} = T_{k}}.
\end{equation}
Each receiver decodes two streams, the common stream and its corresponding private stream.
The common stream is first decoded by treating interference from all private signals as noise.
Successive Interference Cancellation (SIC) is then used to remove the common signal from $y_{k}$, in order to improve the detectability of the private stream.
This is followed by decoding the private stream in the presence of the remaining interference.
The instantaneous (for a given channel state) Signal to Interference plus Noise Ratios (SINRs) of the common stream and the private stream at the output of the $k$th receiver are given by
\begin{equation}
\label{Eq_SINR_MMSE}
  \gamma_{\mathrm{c},k}  \triangleq S_{\mathrm{c},k}I_{\mathrm{c},k}^{-1}  \quad \text{and} \quad
  \gamma_{k}  \triangleq S_{k}I_{k}^{-1}.
\end{equation}
Assuming Gaussian codebooks, the instantaneous achievable information rates for the common stream and the $k$th private stream, from the point of view of the $k$th user, write as
\begin{equation}
\label{Eq_R}
R_{\mathrm{c},k} = \log_{2}(1+\gamma_{\mathrm{c},k}) \quad \text{and} \quad R_{k} =  \log_{2}(1+\gamma_{k}).
\end{equation}
We assume that the transmission is delay-unlimited, and hence channel coding (from messages to symbols) can be preformed over a long sequence of channel states \cite{Tse2005}. This is considered to achieve an average CSIT error performance as we see in the next section.
Precoders on the other hand are adapted throughout the transmission depending on the available estimate of the instantaneous channel state \cite{Caire2007a,Caire2010}. As far as the codewords (symbol streams) are concerned, adaptive precoders are observed as part of the fading channel.
It follows that sending the common message and the $k$th private message at the Ergodic Rates (ERs) given by
$\mathrm{E}_{\mathrm{H}} \left\{R_{\mathrm{c},k} \right\}$ and $\mathrm{E}_{\mathrm{H}} \left\{R_{k} \right\}$ respectively guarantees successful decoding by the $k$th user.
To guarantee that $W_{\mathrm{c}}$ is successfully decoded (and hence cancelled) by all users, it should be transmitted at an ER not exceeding
$\min_{j} \big\{ \mathrm{E}_{\mathrm{H}} \left\{R_{\mathrm{c},j} \right\} \big\}_{j=1}^{K}$.
Finally, the total ER achieved by the $k$th user writes as $\mathrm{E}_{\mathrm{H}} \left\{R_{k} \right\}$ for $k \neq k_{\mathrm{RS}}$, and $\mathrm{E}_{\mathrm{H}} \left\{R_{k} \right\} + \min_{j} \big\{ \mathrm{E}_{\mathrm{H}} \left\{R_{\mathrm{c},j} \right\} \big\}_{j=1}^{K}$ for $k = k_{\mathrm{RS}}$.
%
\section{Motivation, Problem Formulation and DoF Analysis}
\label{Section_Problem_Statement}
In this section, we formulate the precoder design problem for ESR maximization in the RS transmission scheme described in the previous section.
In doing so, we propose the ASR optimization framework which exploits partial instantaneous CSIT to achieve a robust maximized ergodic performance.
Moreover, we derive the DoF performances of the ASR-optimized NoRS and RS designs.
To gain some insight, we start by looking at the conventional (NoRS) ESR maximization problem in the presence of perfect instantaneous CSIT.
This is formulated as
\begin{equation}
\label{Eq_Opt_ESR_Perfect_CSIT}
\max_{\mathrm{E}_{\mathrm{H}}\big\{\mathrm{tr}\big(\mathbf{P}_{\mathrm{p}}\mathbf{P}_{\mathrm{p}}^{H} \big) \big\} \leq P_{\mathrm{t}} }
\mathrm{E}_{\mathrm{H}}\left\{ \sum_{k=1}^{K} R_{k} \right\}.
\end{equation}
The availability of perfect CSIT enables the BS to adapt the precoding matrix $\mathbf{P}_{\mathrm{p}}$ according to $\mathbf{H}$ such that the instantaneous SR (inside the expectation) is maximized.
The ergodic nature of the transmission allows the long-term power constraint in \eqref{Eq_Opt_ESR_Perfect_CSIT}, yielding a form of inter-state (temporal or frequency) power allocation.
A feasible solution for such problem consists of a set of precoding matrices, each for a channel state.
Moreover, we define a precoding scheme as a family of feasible solutions for all possible power levels, i.e.
$\big\{ \mathbf{P}_{\mathrm{p}}(P_{\mathrm{t}},\mathbf{H}) \big\}_{P_{\mathrm{t}},\mathbf{H}}$.
It is common to replace the long-term power constraint in \eqref{Eq_Opt_ESR_Perfect_CSIT} with a short-term one given by
$\mathrm{tr}\big(\mathbf{P}_{\mathrm{p}}\mathbf{P}_{\mathrm{p}}^{H} \big)\leq P_{\mathrm{t}}$ \cite{Love2008,Caire2010}, which corresponds to a precoder peak power constraint in practice.
While this may result in a smaller ESR, it has a significant impact on the tractability of the problem.
In particular, the inter-state power allocation disappears as precoders for different channel states are no longer coupled.
As a result, the maximization in \eqref{Eq_Opt_ESR_Perfect_CSIT} moves inside the expectation, and the optimum solution is obtained by optimizing $\mathbf{P}_{\mathrm{p}}$ separately for each $\mathbf{H}$, i.e.
\begin{equation}
\label{Eq_Opt_SR_Perfect_CSIT}
\max_{\mathrm{tr}\big(\mathbf{P}_{\mathrm{p}}\mathbf{P}_{\mathrm{p}}^{H} \big)  \leq P_{\mathrm{t}} }
\sum_{k=1}^{K} R_{k}
\end{equation}
which corresponds to the problem in \cite{Christensen2008}.
Due to its tractability, we adopt the short-term power constraint.
\subsection{Average Sum Rate Maximization}
\label{Subsection_ASR_Maximization}
Now what if the BS has partial instantaneous CSIT as described in Section \ref{Subsection_CSIT}? The channel state is characterized by $\widehat{\mathbf{H}}$ from the BS's perspective, and it is natural to define a precoding scheme as
$\big\{ \mathbf{P}_{\mathrm{p}}(P_{\mathrm{t}},\widehat{\mathbf{H}}) \big\}_{P_{\mathrm{t}},\widehat{\mathbf{H}}}$.
A naive approach would be to design each precoding matrix by solving problem \eqref{Eq_Opt_SR_Perfect_CSIT} assuming that $\widehat{\mathbf{H}}$ is perfect.
The resulting design is not only unable to cope with the MU interference, it is also unaware of it.
In addition, this may lead to an overestimation of the instantaneous and ergodic rates by the BS, yielding transmission at undecodable rates.
A robust approach on the other hand employs the available CSIT knowledge  to: 1) design an informed precoding scheme that enhances the instantaneous channel condition experienced by the receivers in each channel state, 2) perform transmission at reliable rates and hence guarantee decodability.
In the following, we describe how this can be achieved.
While the BS is unable to predict the instantaneous rates,
it has access to the the Average Rates (ARs) defined as: $\bar{R}_{k} \triangleq \mathrm{E}_{\mathrm{H}\mid\widehat{\mathrm{H}}}\{R_{k} \mid \widehat{\mathbf{H}} \}$ for all $k \in \mathcal{K}$,
determined by the imperfect channel state.
The AR and the ER should not be confused: while the latter describes the long-term performance over all channel states, the former is a short-term (instantaneous) measure that captures the expected performance over the CSIT error distribution for a given channel state estimate.
It turns out that the ERs can be characterized by averaging the ARs over the variation in $\widehat{\mathbf{H}}$.
In particular, for a given fading process and adaptive precoding strategy,
it is evident that
$\mathbf{P}_{\mathrm{p}}(\widehat{\mathbf{H}})$ and
$\bar{R}_{k}\big( \widehat{\mathbf{H}}\big)$ depend on $\widehat{\mathbf{H}}$,
while $R_{k}\big( \mathbf{H},\widehat{\mathbf{H}}\big)$ is determined by $\big\{\mathbf{H},\widehat{\mathbf{H}} \big\}$,
as $\mathbf{P}_{\mathrm{p}}(\widehat{\mathbf{H}})$ is selected by the BS based on the estimate.
From the law of total expectation,
the ER experienced by the user is expressed as
\begin{align}
\nonumber
\mathrm{E}_{\{\mathrm{H},\widehat{\mathrm{H}}\}}\Big\{ R_{k}\big( \mathbf{H},\widehat{\mathbf{H}}\big) \Big\} & =
\mathrm{E}_{\widehat{\mathrm{H}}}\left\{
\mathrm{E}_{\mathrm{H}\mid \widehat{\mathrm{H}}}\big\{ R_{k}\big( \mathbf{H},\widehat{\mathbf{H}} \big) \mid \widehat{\mathbf{H}} \big\} \right\} \\
\label{Eq_ER_EAR}
& = \mathrm{E}_{\widehat{\mathrm{H}}}\left\{ \bar{R}_{k}\big( \widehat{\mathbf{H}}\big)  \right\}
\end{align}
from which the ESR for a given precoding strategy writes as $\mathrm{E}_{\widehat{\mathrm{H}}}\left\{ \sum_{k=1}^{K} \bar{R}_{k} \right\}$.
Following the same logic in \eqref{Eq_Opt_ESR_Perfect_CSIT} and \eqref{Eq_Opt_SR_Perfect_CSIT},
maximizing the ESR under partial CSIT and a short-term power constraint is achieved by optimizing $\mathbf{P}_{\mathrm{p}}\big(\widehat{\mathbf{H}}\big)$  such that the ASR,
i.e. $\sum_{k=1}^{K}\bar{R}_{k}$, is maximized for each $\widehat{\mathbf{H}}$.
For a class of information theoretic channels under imperfect CSIT with a Markov property,
it has been shown that the ergodic capacity is found in a similar manner,
while replacing the ASR with the average mutual-information, and maximizing
over the input distribution rather than the precoding matrix \cite{Caire2007a}.
This gives rise to the ASR optimization problem formulated as
\begin{equation}
\label{Eq_Opt_ASR_NoRS}
\mathcal{R}(P_{\mathrm{t}}):
\begin{cases}
       \underset{\mathbf{P}_{\mathrm{p}} }{\max} & \sum_{k=1}^{K} \bar{R}_{k}  \\
       \text{s.t.}  & \mathrm{tr}\big(\mathbf{P}_{\mathrm{p}}\mathbf{P}_{\mathrm{p}}^{H}\big) \leq P_{\mathrm{t}}.
\end{cases}
\end{equation}
It follows that the maximized ESR given by $\mathrm{E}_{\widehat{\mathrm{H}}}\left\{ \mathcal{R}(P_{\mathrm{t}}) \right\}$ is achievable.
The fixed-coding/adaptive-precoding strategy
described in Section \ref{Subsection_SINR_Rate} can be employed, where the knowledge of the long-term properties of CSIT is leveraged to predict the ERs and adjust the  code rates, while precoders are adapted each time a new channel estimate is revealed.
It is evident that for perfect CSIT, \eqref{Eq_Opt_ASR_NoRS} reduces to the SR problem in \eqref{Eq_Opt_SR_Perfect_CSIT}.
A discussion on the influence of CSIT uncertainty on the ESR performance follows in the next subsection.
It is worth noting that the considered CSIT model reduces to the one in \cite{Negro2012,Bashar2014}
when $\widehat{\mathbf{H}}$ is fixed over all channel states, and the fast-fading model in \cite{Razaviyayn2013a} when
$\widehat{\mathbf{H}}$ is completely suppressed.
Both lead to special cases of the problem formulation, where the ESR problem and the ASR problem coincide,
and a precoding scheme is given by a single precoding matrix for each $P_{\mathrm{t}}$.
Now we turn to designing a precoding scheme
$\big\{ \mathbf{P}(P_{\mathrm{t}},\widehat{\mathbf{H}}) \big\}_{P_{\mathrm{t}},\widehat{\mathbf{H}}}$ that maximizes the ESR performance of the RS strategy by formulating the corresponding ASR problem.
The the $k$th common AR is defined as $\bar{R}_{\mathrm{c},k}
\triangleq \mathrm{E}_{\mathrm{H} \mid \widehat{\mathrm{H}}}\big\{R_{\mathrm{c},k} \mid \widehat{\mathbf{H}} \big\}$,
from which the common AR is given by $\bar{R}_{\mathrm{c}} \triangleq \min_{j} \{ \bar{R}_{\mathrm{c},j} \}_{j=1}^{K}$.
The ASR is expressed as
$\bar{R}_{\mathrm{c}} + \sum_{k=1}^{K} \bar{R}_{k}$, and the ASR maximization problem is formulated as
\begin{equation}
\label{Eq_Opt_ASR_RS}
\mathcal{R}_{\mathrm{RS}}(P_{\mathrm{t}}):
\begin{cases}
       \underset{\bar{R}_{c}, \mathbf{P} }{\max} &
 \bar{R}_{\mathrm{c}}+ \sum_{k=1}^{K} \bar{R}_{k}  \\
       \text{s.t.}  & \bar{R}_{\mathrm{c},k} \geq \bar{R}_{\mathrm{c}}, \; \forall k\in\mathcal{K} \\
                    & \mathrm{tr}\big(\mathbf{P}\mathbf{P}^{H}\big) \leq P_{\mathrm{t}}
\end{cases}
\end{equation}
where the inequality constraints involving $\bar{R}_{\mathrm{c}}$ are equivalent to the pointwise minimization in the definition of $\bar{R}_{\mathrm{c}}$.
Next, we show that the ESR given by $\mathrm{E}_{\widehat{\mathrm{H}}}\left\{ \mathcal{R}_{\mathrm{RS}}(P_{\mathrm{t}}) \right\}$ is achievable.
The reliability of the private ERs follows from \eqref{Eq_ER_EAR}.
As for the common ER, we write
\begin{equation}
\label{Eq_ERc_EARc}
\min_{j\in \mathcal{K}} \Big\{ \mathrm{E}_{\{\mathrm{H},\widehat{\mathrm{H}}\}}\big\{ R_{\mathrm{c},j} \big\} \Big\} =
\min_{j\in \mathcal{K}} \Big\{ \mathrm{E}_{\widehat{\mathrm{H}}}\big\{ \bar{R}_{\mathrm{c},j} \big\} \Big\} \geq
 \mathrm{E}_{\widehat{\mathrm{H}}}\Big\{ \min_{j\in \mathcal{K}} \bar{R}_{\mathrm{c},j} \Big\}
\end{equation}
where the equality follows from the law of total expectation as in \eqref{Eq_ER_EAR}, and the inequality follows from the fact that
moving the minimization inside the expectation does not increase the value.
Leaving the minimization outside the expectation couples the common ARs, and hence $\mathbf{P}\big(\widehat{\mathbf{H}}\big)$, across the different channel states, leading to an intractable formulation.
After formulating the NoRS and RS instantaneous ASR problems, the next two questions that come to mind are: 1) how does the instantaneous CSIT quality influence the long-term performances given by
$\mathrm{E}_{\widehat{\mathrm{H}}}\left\{ \mathcal{R}(P_{\mathrm{t}}) \right\}$ and
$\mathrm{E}_{\widehat{\mathrm{H}}}\left\{ \mathcal{R}_{\mathrm{RS}}(P_{\mathrm{t}}) \right\}$?
and 2) How do the NoRS and RS performances compare?
To answer these question, we resort to DoF analysis.
\subsection{DoF Performance}
The DoFs for the NoRS and RS strategies are defined as
\begin{equation}
\label{Eq_Optimum_DoF}
\lim_{P_{\mathrm{t}}\rightarrow\infty}
\frac{ \mathrm{E}_{\widehat{\mathrm{H}}}\left\{ \mathcal{R}(P_{\mathrm{t}}) \right\} }{\log_{2}(P_{\mathrm{t}})}
\quad \text{and} \quad
\lim_{P_{\mathrm{t}}\rightarrow\infty} \frac{ \mathrm{E}_{\widehat{\mathrm{H}}}\left\{ \mathcal{R}_{\mathrm{RS}}(P_{\mathrm{t}}) \right\} }{\log_{2}(P_{\mathrm{t}})}.
\end{equation}
The DoF can be roughly interpreted as the total number of interference-free streams that can be simultaneously supported in a single channel use. This follows by noting that the rate of a single interference-free data stream scales as
$\log_{2}(P_{\mathrm{t}}) + O(1)$.
The significance of the DoF in such setups comes from the detrimental effects MU interference may have, and the role of CSIT in dealing with such effect.
For the proof of our next result, we assume isotropically distributed CSIT error vectors, i.e.
$\mathbf{R}_{\mathrm{e},k} = \frac{\sigma_{\mathrm{e},k}^{2}}{N_{\mathrm{t}}} \mathbf{I}$ for all $k \in \mathcal{K}$.
Moreover, we assume that $\widehat{\mathbf{H}}$ is of full column rank with probability one.
Note that these assumptions are not necessary for the optimization in the following sections.
\newtheorem{Theorem_RS_DoF}[Theorem_Counter]{Theorem}
\begin{Theorem_RS_DoF}\label{Theorem_RS_DoF}
\textnormal{The DoF of an optimum NoRS scheme is
\begin{equation}
\label{Eq_DoF_no_RS}
\lim_{P_{\mathrm{t}}\rightarrow\infty} \frac{ \mathrm{E}_{\widehat{\mathrm{H}}}\left\{ \mathcal{R}(P_{\mathrm{t}}) \right\} }{\log_{2}(P_{\mathrm{t}})}
= \max{\{1,K \alpha\}}
\end{equation}
while the DoF of an optimum RS scheme is given by
\begin{equation}
\label{Eq_DoF_RS}
\lim_{P_{\mathrm{t}}\rightarrow\infty} \frac{ \mathrm{E}_{\widehat{\mathrm{H}}}\left\{ \mathcal{R}_{\mathrm{RS}}(P_{\mathrm{t}}) \right\} }{\log_{2}(P_{\mathrm{t}})}
= 1 + (K-1) \alpha.
\end{equation}
%
}
\end{Theorem_RS_DoF}
%
The proof is relegated to Appendix \ref{Appendix_Proof_Theorem_RS_DoF}.
It is evident that when $\alpha = 1$, the maximum attainable DoF is achieved, i.e. $K$.
Although CSIT may still be erroneous, it is good enough to design private precoders that reduce the MU interference to the level of additive noise.
However, this is not possible when $\alpha$ drops below $1$, as private streams start leaking substantial interference.
The NoRS scheme exhibits loss in DoF until $\alpha = 1/K$. As $\alpha$ drops below $1/K$, the CSIT quality is not good enough to support multi-stream transmission, and it is preferred from a DoF perspective to switch to SU transmission by allocating all power to one stream, hence achieving a DoF of 1.
In naive designs with fixed and uniform power allocation across users \cite{Jindal2006,Caire2010}, the DoF reduces to zero for $\alpha = 0$, and the gain from switching to SU transmission is not realized.
Now looking at RS, as soon as $\alpha$ drops below $1$, the scheme carefully allocates powers to the private streams such that MU interference is always received at the level of additive noise. This is achieved by scaling down private powers to $O(P_{\mathrm{t}}^{\alpha})$, hence maintaining a DoF of $K\alpha$.
The rest of the power, which scales as $O(P_{\mathrm{t}})$, is allocated to the common stream which is broadcasted to all users.
This achieves a DoF gain of $1-\alpha$, as interference from private messages is treated as noise.
It should be noted that the RS DoF is strictly greater than the NoRS DoF for all $\alpha \in (0,1)$.

The RS DoF in \eqref{Eq_DoF_RS} is inline with the results in \cite{Yang2013,Hao2013}. Although optimizing the precoders does not improve the achievable DoF in these works, simulation results in Section \ref{Section_Numerical_Results} show that optimized RS schemes are superior from a rate performance perspective.
It is worth highlighting that the converse result recently reported in \cite{Davoodi2016} proves the optimality of the DoF in \eqref{Eq_DoF_RS} for the Gaussian MISO-BC when one user has perfect CSIT.
Since improving the CSIT quality of one user cannot decrease the DoF, the optimality of the result in \eqref{Eq_DoF_RS} naturally follows.
We conclude this section by highlighting that further to the high SNR analysis, $\mathcal{R}_{\mathrm{RS}}(P_{\mathrm{t}}) \geq \mathcal{R}(P_{\mathrm{t}})$ is guaranteed over the entire range of SNRs. This is seen by noting that solving
\eqref{Eq_Opt_ASR_NoRS} is equivalent to solving \eqref{Eq_Opt_ASR_RS} over a subset of its domain characterized by restricting $\|\mathbf{p}_{\mathrm{c}} \| ^{2}$ to zero.
In the following, we focus on solving the RS design problem, achieved by solving \eqref{Eq_Opt_ASR_RS}.
\section{Sample Average Approximated WMMSE Algorithm}
\label{Section_ASR_AWSMSE}
The RS ASR problem in \eqref{Eq_Opt_ASR_RS} is of a stochastic nature and appears to be challenging to solve in its current form. In fact, deterministic subproblems of \eqref{Eq_Opt_ASR_RS} are known to be non-convex and non-trivial \cite{Christensen2008,Sidiropoulos2006}.
We propose a two-step approach to obtain a tractable form of \eqref{Eq_Opt_ASR_RS}.
In the first step, a deterministic approximation is obtained using the SAA method \cite{Shapiro2009}.
In the second step, the deterministic ASR problem is transformed into an equivalent, and solvable, augmented Average Weighted Sum MSE (AWSMSE) problem.
Before we proceed, it should be highlighted that in the following analysis, the dependencies of some functions on certain relevant variables is occasionally emphasized in the notation.
\subsection{Sample Average Approximation}
\label{Subsection_SAA}
For a given estimate $\widehat{\mathbf{H}}$ and index set $\mathcal{M} \triangleq \{1, \ldots, M \}$,
let 
\begin{equation}
\nonumber
\mathbb{H}^{(M)} \triangleq \big\{ \mathbf{H}^{(m)} = \widehat{\mathbf{H}} + \widetilde{\mathbf{H}}^{(m)} \mid \widehat{\mathbf{H}}, \ m \in \mathcal{M} \big\}
\end{equation}
be a sample of $M$ i.i.d realizations drawn from the conditional distribution with density
$f_{\mathrm{H} \mid \widehat{\mathrm{H}}} \big( \mathbf{H} \mid \widehat{\mathbf{H}} \big)$.
This sample is used to approximate the ARs through their Sample Average Functions (SAFs) defined as:
$\bar{R}_{\mathrm{c},k}^{(M)} \triangleq \frac{1}{M} \sum_{m=1}^{M} R_{\mathrm{c},k}^{(m)} $ and $\bar{R}_{k}^{(M)} \triangleq  \frac{1}{M} \sum_{m=1}^{M} R_{k}^{(m)}$, where
$R_{\mathrm{c},k}^{(m)} \triangleq R_{\mathrm{c},k}\big(\mathbf{H}^{(m)} \big)$
and
$R_{k}^{(m)} \triangleq  R_{k}(\mathbf{H}^{(m)}\big)$
are rates associated with the $m$th realization.
This leads to the SAA of problem \eqref{Eq_Opt_ASR_RS} formulated as
%
%
\begin{equation}
 \label{Eq_Opt_ASR_RS_M}
\mathcal{R}_{\mathrm{RS}}^{(M)}(P_{\mathrm{t}}):
\begin{cases}
       \underset{\bar{R}_{\mathrm{c}}, \mathbf{P} }{\max} &
 \bar{R}_{\mathrm{c}}+ \sum_{k=1}^{K} \bar{R}_{k}^{(M)}  \\
       \text{s.t.}  & \bar{R}_{\mathrm{c},k}^{(M)} \geq \bar{R}_{\mathrm{c}}, \; \forall k\in\mathcal{K} \\
                    & \mathrm{tr}\big(\mathbf{P}\mathbf{P}^{H}\big) \leq P_{\mathrm{t}}.
\end{cases}
\end{equation}
%
It should be noted that $\mathbf{P}$ is fixed over the $M$ realizations, which follows from the definition of the ARs.
Before we proceed, we make the following finite SNR/SINR assumption.
\newtheorem{Assumption_Bounded_SNR}[Assumption_Counter]{Assumption}
\begin{Assumption_Bounded_SNR}
\label{Assumption_Bounded_SNR}
\textnormal{
In the following we assume that $\mathrm{SNR} < \infty$, i.e. $\sigma_{\mathrm{n}}^{2} >0$ and $P_{\mathrm{t}} < \infty$.
Moreover, we assume that channel realizations are bounded.
Hence, $\gamma_{\mathrm{c},k},\gamma_{k} < \infty$ for all $k \in \mathcal {K}$.
}
\end{Assumption_Bounded_SNR}
%
It follows that all rates are bounded for any $P_{\mathrm{t}}$ and all channel realizations. Therefore, from the strong Law of Large Numbers (LLN), we have
\begin{subequations}
\label{Eq_R_LLN}
\begin{align}
   \lim_{M\rightarrow \infty} \bar{R}_{\mathrm{c},k}^{(M)}(\mathbf{P}) & = \bar{R}_{\mathrm{c},k}(\mathbf{P}), 
   \ \text{a.s.} \ \forall \mathbf{P} \in \mathbb{P} \\
   \lim_{M\rightarrow \infty} \bar{R}_{k}^{(M)}(\mathbf{P}) & = \bar{R}_{k}(\mathbf{P}), \ \text{a.s.} \ \forall \mathbf{P} \in \mathbb{P}
\end{align}
\end{subequations}
where the dependency on $\mathbf{P}$ is highlighted, a.s. denotes almost surely, and $\mathbb{P} \triangleq \big\{ \mathbf{P} \mid  \mathrm{tr}\big(\mathbf{P}\mathbf{P}^{H}\big) \leq P_{\mathrm{t}} \big\}$ is the feasible set of precoders.
\eqref{Eq_R_LLN} suggests that the optimum solution of problem \eqref{Eq_Opt_ASR_RS_M} converges to its counterpart of problem \eqref{Eq_Opt_ASR_RS} as $M \rightarrow \infty$. In the following, we show that this is indeed the case.

First, we note that $\mathbb{P}$ is compact and the rate functions are bounded and continuously differentiable in $\mathbf{P}$.
It follows that the convergence in \eqref{Eq_R_LLN} is uniform in $\mathbf{P} \in \mathbb{P}$.
The ARs are also continuously differentiable with gradients given by $\nabla_{\mathbf{P}} \bar{R}_{\mathrm{c},k}(\mathbf{P}) =  \mathrm{E}_{\mathrm{H} \mid \widehat{\mathrm{H}}} \big\{\nabla_{\mathbf{P}} R_{\mathrm{c},k}(\mathbf{P}) \mid \widehat{\mathbf{H}} \big\}$
and
$\nabla_{\mathbf{P}} \bar{R}_{k}(\mathbf{P}) =  \mathrm{E}_{\mathrm{H} \mid \widehat{\mathrm{H}}} \big\{\nabla_{\mathbf{P}}R_{k}(\mathbf{P}) \mid \widehat{\mathbf{H}} \big\}$,
which follows from the bounded convergence theorem \cite{Fristedt1997}.
On the other hand, the objective functions of \eqref{Eq_Opt_ASR_RS} and \eqref{Eq_Opt_ASR_RS_M} can be equivalently reformulated by incorporating the common rates as
\begin{subequations}
\label{Eq_R_s_M}
\begin{align}
\bar{R}_{\mathrm{s}}(\mathbf{P}) & \triangleq \min_{j} \big\{\bar{R}_{\mathrm{c},j}(\mathbf{P}) \big\}_{j=1}^{K} + \sum_{k=1}^{k} \bar{R}_{k}(\mathbf{P}) \\
\bar{R}_{\mathrm{s}}^{(M)}(\mathbf{P}) & \triangleq \min_{j} \big\{ \bar{R}_{\mathrm{c},j}^{(M)} (\mathbf{P}) \big\}_{j=1}^{K}  + \sum_{k=1}^{k} \bar{R}_{k}^{(M)}(\mathbf{P})
\end{align}
\end{subequations}
where $\bar{R}_{\mathrm{s}}(\mathbf{P})$ and $\bar{R}_{\mathrm{s}}^{(M)}(\mathbf{P})$ are the equivalent ASRs.
Hence, from \eqref{Eq_R_LLN}, \eqref{Eq_R_s_M} and the continuity of the pointwise minimization, we have
\begin{equation}
\label{Eq_R_s_LLN}
\lim_{M \rightarrow \infty} \bar{R}_{\mathrm{s}}^{(M)}(\mathbf{P}) =
\bar{R}_{\mathrm{s}}(\mathbf{P}), \ \text{a.s.} \ \forall \mathbf{P} \in \mathbb{P}.
\end{equation}
From the continuity of the ARs, we observe that $\bar{R}_{\mathrm{s}}(\mathbf{P})$ is continuous in $\mathbf{P}$ although not necessarily differentiable at all points due to the embedded pointwise minimization.
Hence, it can be shown that the convergence in \eqref{Eq_R_s_LLN} is also uniform.
Combining these observations with \cite[Theorem 5.3]{Shapiro2009}, it is concluded that
the set of global optimum solutions of the SAA problem in \eqref{Eq_Opt_ASR_RS_M} converges to that of the stochastic problem in \eqref{Eq_Opt_ASR_RS} a.s. as $M \rightarrow \infty$.
Extending this result to the set of points that satisfy the first-order optimality conditions (i.e. KKT points) can be found in \cite[Section 4.3]{Kim2015}.
Now, we turn to solving the sampled problem in \eqref{Eq_Opt_ASR_RS_M}.
%
\subsection{Augmented AWSMSE Minimization}
\label{Subsection_Augmented_AWSMSE}
Although \eqref{Eq_Opt_ASR_RS_M} is deterministic, it is non-convex and  very challenging to solve.
Therefore, the approach proposed in \cite{Christensen2008} is employed to reformulate \eqref{Eq_Opt_ASR_RS_M} into an equivalent augmented WMSE form.
Augmented WMSE problems differ from their conventional WMSE counterparts (e.g. \cite{Shenouda2008}) in two ways: 1) the weights are considered as optimization variables, 2) the cost function is extended to incorporate the logarithms of the weights.
A Rate-WMMSE relationship is built upon the two aforementioned features.
Let $\widehat{s}_{\mathrm{c},k}=g_{\mathrm{c},k} y_{k}$ be $k$th user's estimate of $s_{\mathrm{c}}$,  where $g_{\mathrm{c},k} $ is a scalar equalizer.
After successfully removing the common stream, the estimate of $s_{k}$ is obtained as $\widehat{s}_{k}=g_{k}(y_{k}-\mathbf{h}_{k}^{H}\mathbf{p}_{\mathrm{c}}s_{\mathrm{c},k})$,
where $g_{k}$ is the corresponding equalizer.
At the output of the $k$th receiver, the common and private MSEs are defined as $\varepsilon_{\mathrm{c},k} \triangleq \mathrm{E}\{|\widehat{s}_{\mathrm{c},k} - s_{\mathrm{c}}|^{2}\}$ and $\varepsilon_{k} \triangleq \mathrm{E}\{|\widehat{s}_{k} - s_{k}|^{2}\}$ respectively, which write as:
\begin{subequations}
\label{Eq_MSE}
\begin{align}
  \label{Eq_MSE_c_k}
  \varepsilon_{\mathrm{c},k} & = |g_{\mathrm{c},k}|^{2} T_{\mathrm{c},k} -2\Re \big\{g_{\mathrm{c},k}\mathbf{h}_{k}^{H}\mathbf{p}_{\mathrm{c}}\big\}+1 \\
  \label{Eq_MSE_k}
  \varepsilon_{k} & =  |g_{k}|^{2} T_{k}-2\Re \big\{g_{k}\mathbf{h}_{k}^{H}\mathbf{p}_{k}\big\}+1
\end{align}
\end{subequations}
where $T_{\mathrm{c},k}$ and  $T_{k}$ are defined in \eqref{Eq_T_c_k}.
Optimum Minimum MSE (MMSE) equalizers are given as
\begin{equation}
 \label{Eq_g_MMSE}
  g_{\mathrm{c},k}^{\mathrm{MMSE}} = \mathbf{p}_{\mathrm{c}}^{H}\mathbf{h}_{k} T_{\mathrm{c},k}^{-1}
  \quad \text{and} \quad
  g_{k}^{\mathrm{MMSE}} = \mathbf{p}_{k}^{H}\mathbf{h}_{k}T_{k}^{-1}
\end{equation}
obtained by solving $\frac{\partial \varepsilon_{\mathrm{c},k}}{\partial g_{\mathrm{c},k}}= 0$ and 
$\frac{\partial \varepsilon_{k}}{\partial g_{k}} = 0$. Substituting \eqref{Eq_g_MMSE} into \eqref{Eq_MSE}, the MMSEs write as
\begin{subequations}
\label{Eq_MMSE}
\begin{align}
\varepsilon_{\mathrm{c},k}^{\mathrm{MMSE}} & \triangleq \underset{g_{\mathrm{c},k}}{\min} \ \varepsilon_{\mathrm{c},k} =  T_{\mathrm{c},k}^{-1} I_{\mathrm{c},k} \\
\varepsilon_{k}^{\mathrm{MMSE}} & \triangleq \underset{g_{k}}{\min} \ \varepsilon_{k} = T_{k}^{-1}I_{k}
\end{align}
\end{subequations}
from which the SINRs are rewritten as $ \gamma_{\mathrm{c},k}  = \big( 1/\varepsilon_{\mathrm{c},k}^{\mathrm{MMSE}} \big) - 1$  and
$  \gamma_{k} =\big(1/\varepsilon_{k}^{\mathrm{MMSE}}\big) - 1$, and the rates write as
$R_{\mathrm{c},k} =-\log_{2}(\varepsilon_{\mathrm{c},k}^{\mathrm{MMSE}})$
and
$R_{k} =-\log_{2}(\varepsilon_{k}^{\mathrm{MMSE}})$.
Now, the augmented WMSEs are given by
\begin{equation}
\label{Eq_A_WMSEs}
\xi_{\mathrm{c},k}
 =
u_{\mathrm{c},k} \varepsilon_{\mathrm{c},k}  -  \log_{2} (  u_{\mathrm{c},k} )
\ \ \text{and} \ \
\xi_{k}
 =
u_{k} \varepsilon_{k}  -  \log_{2}  (  u_{k} )
\end{equation}
where $u_{\mathrm{c},k}, u_{k} > 0$ are weights associated with the $k$th user's MSEs.
In the following, $\xi_{\mathrm{c},k}$ and $\xi_{k}$ are referred to as  the WMSEs, where "augmented" is dropped for brevity.
By taking the equalizers and weights as optimization variables, the Rate-WMMSE relationship is established as
\begin{subequations}
\label{Eq_min_WMSE}
\begin{align}
 \xi_{\mathrm{c},k}^{\mathrm{MMSE}} & \triangleq \underset{u_{\mathrm{c},k}, g_{\mathrm{c},k}}{\min} \xi_{\mathrm{c},k} = 1-R_{\mathrm{c},k} \\
 \xi_{k}^{\mathrm{MMSE}} & \triangleq \underset{u_{k}, g_{k}}{\min} \ \xi_{k}= 1-R_{k}
\end{align}
\end{subequations}
which is obtained as follows. From
$\frac{\partial\xi_{\mathrm{c},k}}{\partial g_{\mathrm{c},k}} = 0$
and
$\frac{\partial\xi_{k}}{\partial g_{k}} = 0$, the optimum equalizers are given as
$g_{\mathrm{c},k}^{\ast}  = g_{\mathrm{c},k}^{\mathrm{MMSE}} $ and $g_{k}^{\ast} = g_{k}^{\mathrm{MMSE}}$.
Substituting this back into \eqref{Eq_A_WMSEs} yields
\begin{subequations}
\label{Eq_A_WMMSEs}
\begin{align}
\xi_{\mathrm{c},k}\big(g_{\mathrm{c},k}^{\mathrm{MMSE}}\big) & =   u_{\mathrm{c},k}\varepsilon_{\mathrm{c},k}^{\mathrm{MMSE}} - \log_{2}(u_{\mathrm{c},k})
\\
\xi_{k}\big(g_{k}^{\mathrm{MMSE}}\big) & =   u_{k}\varepsilon_{k}^{\mathrm{MMSE}} - \log_{2}(u_{k}).
\end{align}
\end{subequations}
Furthermore, from
$\frac{\partial \xi_{\mathrm{c},k}(g_{\mathrm{c},k}^{\mathrm{MMSE}})}{\partial u_{\mathrm{c},k}} = 0$
and
$\frac{\partial \xi_{k}(g_{k}^{\mathrm{MMSE}})}{\partial u_{k}} = 0$, we obtain the optimum MMSE weights given by
$u_{\mathrm{c},k}^{\ast} = u_{\mathrm{c},k}^{\mathrm{MMSE}} \triangleq \big( \varepsilon_{\mathrm{c},k}^{\mathrm{MMSE}} \big)^{-1}$
and
$u_{k}^{\ast} = u_{k}^{\mathrm{MMSE}} \triangleq \big( \varepsilon_{k}^{\mathrm{MMSE}} \big)^{-1}$,
where a scaling factor of $\big( \ln(2)\big)^{-1}$ has been omitted as it has no effect on the solution.
Substituting this back into \eqref{Eq_A_WMMSEs} yields the relationship in \eqref{Eq_min_WMSE}.

By taking the expectation over the conditional distribution of $\mathbf{H}$ given $\widehat{\mathbf{H}}$, an average version of the Rate-WMMSE relationship, denoted by AR-AWMMSE, is expressed as
\begin{subequations}
\label{Eq_min_AWMSE}
\begin{align}
\bar{\xi}_{\mathrm{c},k}^{\mathrm{MMSE}} & \triangleq  \mathrm{E}_{\mathrm{H}\mid \widehat{\mathrm{H}}}
\left\{ \underset{u_{\mathrm{c},k}, g_{\mathrm{c},k}}{\min}\xi_{\mathrm{c},k} \mid \widehat{\mathbf{H}}  \right\} = 1-\bar{R}_{\mathrm{c},k}
 \\
\bar{\xi}_{k}^{\mathrm{MMSE}} & \triangleq  \mathrm{E}_{\mathrm{H}\mid \widehat{\mathrm{H}}}
\left\{\underset{u_{k}, g_{k}}{\min} \; \xi_{k} \mid \widehat{\mathbf{H}} \right\}= 1-\bar{R}_{k}
\end{align}
\end{subequations}
where the expectations are taken outside the minimizations to account for the dependencies of the optimum equalizers and weights on
$\mathbf{H}$.
The rational behind employing the Rate-WMMSE relationship is explained as follows. The incorporation of the equalizers and weights unveils a block-wise convexity property, which can be checked by noting that the WMSEs are convex in each of their corresponding variables when all other variables are fixed. This will prove crucial for solving the problem.

To formulate a deterministic equivalent of \eqref{Eq_min_AWMSE}, the Average WMSEs (AWMSEs) are approximated by their SAFs.
The sampling extends to the equalizers and weights such that
%
\begin{equation}
\label{Eq_AWMSE_M}
\bar{\xi}_{\mathrm{c},k}^{(M)} \triangleq \frac{1}{M} \sum_{m=1}^{M}\xi_{\mathrm{c},k}^{(m)}
\quad \text{and} \quad
\bar{\xi}_{k}^{(M)} \triangleq \frac{1}{M} \sum_{m=1}^{M}\xi_{k}^{(m)}
\end{equation}
where $\xi_{\mathrm{c},k}^{(m)}  \triangleq \xi_{\mathrm{c},k}\big(\mathbf{h}_{k}^{(m)},g_{\mathrm{c},k}^{(m)},u_{\mathrm{c},k}^{(m)}\big)$
and
$\xi_{k}^{(m)}  \triangleq \xi_{k}\big(\mathbf{h}_{k}^{(m)},g_{k}^{(m)},u_{k}^{(m)}\big)$,
$g_{\mathrm{c},k}^{(m)}  \triangleq   g_{\mathrm{c},k}\big(  \mathbf{h}_{k}^{(m)} \big)  $
and
$g_{k}^{(m)} \triangleq   g_{k}\big( \mathbf{h}_{k}^{(m)} \big)  $, and
$u_{\mathrm{c},k}^{(m)}   \triangleq u_{\mathrm{c},k}\big(   \mathbf{h}_{k}^{(m)} \big)  $
and
$u_{k}^{(m)}  \triangleq  u_{k}\big( \mathbf{h}_{k}^{(m)} \big)  $, are all associated with the $m$th realization in $\mathbb{H}^{(M)}$.
For compactness, we define the set of sampled equalizers as:
$\mathbf{G}  \triangleq   \big\{  \mathbf{g}_{\mathrm{c},k},\mathbf{g}_{k}  \mid   k  \in   \mathcal{K}   \big\}$,
where
$\mathbf{g}_{\mathrm{c},k}  \triangleq   \big\{   g_{\mathrm{c},k}^{(m)}   \mid   m   \in   \mathcal{M}   \big\}$
and
$\mathbf{g}_{k}  \triangleq   \big\{  g_{k}^{(m)}   \mid   m   \in  \mathcal{M}   \big\}$.
In a similar manner, we define:
$\mathbf{U}   \triangleq   \big\{   \mathbf{u}_{\mathrm{c},k},\mathbf{u}_{k}   \mid   k  \in   \mathcal{K}  \big\}$,
where
$\mathbf{u}_{\mathrm{c},k}   \triangleq   \big\{   u_{\mathrm{c},k}^{(m)}   \mid   m   \in   \mathcal{M}   \big\}$
and
$\mathbf{u}_{k}   \triangleq  \big\{   u_{k}^{(m)}   \mid   m  \in   \mathcal{M}  \big\}$.
The same approach used to prove \eqref{Eq_min_WMSE} is employed to demonstrate the following relationship
\begin{subequations}
\label{Eq_min_AWMSE_M}
\begin{align}
\bar{\xi}_{\mathrm{c},k}^{\mathrm{MMSE}(M)} \triangleq
\underset{\mathbf{u}_{\mathrm{c},k}, \mathbf{g}_{\mathrm{c},k}}{\min} \bar{\xi}_{\mathrm{c},k}^{(M)} = 1-\bar{R}_{\mathrm{c},k}^{(M)}
\\
\bar{\xi}_{k}^{\mathrm{MMSE}(M)} \triangleq
\underset{\mathbf{u}_{k}, \mathbf{g}_{k}}{\min} \ \bar{\xi}_{k}^{(M)}= 1-\bar{R}_{k}^{(M)}
 \end{align}
\end{subequations}
where optimality conditions are checked separately for each conditional  realization.
The sets of optimum MMSE equalizers associated with \eqref{Eq_min_AWMSE_M} are defined as
$\mathbf{g}^{\mathrm{MMSE}}_{\mathrm{c},k}  \triangleq  \big\{  g_{\mathrm{c},k}^{\mathrm{MMSE}(m)}   \mid  m  \in   \mathcal{M}  \big\}$
and
$\mathbf{g}^{\mathrm{MMSE}}_{k}  \triangleq  \big\{  g_{k}^{\mathrm{MMSE}(m)}  \mid m   \in   \mathcal{M}   \big\}$.
In the same manner, the sets of optimum MMSE weights are defined as
$\mathbf{u}^{\mathrm{MMSE}}_{\mathrm{c},k}  \triangleq  \big\{  u_{\mathrm{c},k}^{\mathrm{MMSE}(m)}  \mid  m   \in   \mathcal{M}  \big\}$
and
$\mathbf{u}^{\mathrm{MMSE}}_{k}  \triangleq  \big\{  u_{k}^{\mathrm{MMSE}(m)}  \mid  m   \in   \mathcal{M}  \big\}$.
For the $K$ users, the MMSE solution is composed as
$\mathbf{G}^{\mathrm{MMSE}}  \triangleq   \big\{  \mathbf{g}_{\mathrm{c},k}^{\mathrm{MMSE}}, \mathbf{g}_{k}^{\mathrm{MMSE}}  \mid  k   \in  \mathcal{K}  \big\}$
and
$\mathbf{U}^{\mathrm{MMSE}}  \triangleq   \big\{  \mathbf{u}_{\mathrm{c},k}^{\mathrm{MMSE}}, \mathbf{u}_{k}^{\mathrm{MMSE}}  \mid  k  \in  \mathcal{K}  \big\}$.
Motivated by the relationship in \eqref{Eq_min_AWMSE_M}, the deterministic augmented AWSMSE minimization problem is formulated as
\begin{equation}
 \label{Eq_Opt_AWSMSE_M}
\mathcal{A}_{\mathrm{RS}}^{(M)}(P_{\mathrm{t}}):
\begin{cases}
       \underset{\bar{\xi}_{\mathrm{c}}, \mathbf{P}, \mathbf{U}, \mathbf{G} }{\min} &
 \bar{\xi}_{\mathrm{c}}+ \sum_{k=1}^{K} \bar{\xi}_{k}^{(M)}  \\
      \ \ \text{s.t.}  & \bar{\xi}_{\mathrm{c},k}^{(M)} \leq \bar{\xi}_{\mathrm{c}}, \; \forall k\in\mathcal{K} \\
                    & \mathrm{tr}\big(\mathbf{P}\mathbf{P}^{H}\big) \leq P_{\mathrm{t}}
\end{cases}
\end{equation}
%
where $\bar{\xi}_{\mathrm{c}}$ represents the common AWMSE.
\subsection{Equivalence}
\label{Subsection_AWMSE_Rare_Equivalence}
The fact that the AWMSEs are decoupled in their corresponding equalizers and weights suggests that optimizing \eqref{Eq_Opt_AWSMSE_M} w.r.t $(\mathbf{U},\mathbf{G} )$  is achieved by minimizing each of the AWMSEs individually as shown in \eqref{Eq_min_AWMSE_M}.
This can be confirmed by showing that the MMSE solution $\big(\mathbf{U}^{\mathrm{MMSE}},\mathbf{G}^{\mathrm{MMSE}}\big)$ satisfies the KKT optimality conditions of \eqref{Eq_Opt_AWSMSE_M} for a given $\mathbf{P}$.
As a result, it is easily shown that under the MMSE solution, \eqref{Eq_Opt_AWSMSE_M} boils down to \eqref{Eq_Opt_ASR_RS_M} with affine transformations applied to the objective function and the common rate, i.e. $\mathcal{A}_{\mathrm{RS}}^{(M)}(P_{\mathrm{t}}) = K+ 1 - \mathcal{R}_{\mathrm{RS}}^{(M)}(P_{\mathrm{t}})$ and $\bar{\xi}_{\mathrm{c}} = 1 - \bar{R}_{\mathrm{c}}$.
This Rate-WMMSE equivalence is not restricted to the global optimum solutions, and in fact can be extended to the whole set of stationary points.
In particular, for any point $\left(\bar{\xi}_{\mathrm{c}}^{\ast},\mathbf{P}^{\ast},\mathbf{U}^{\ast},\mathbf{G}^{\ast}\right)$ satisfying the KKT optimality conditions of \eqref{Eq_Opt_AWSMSE_M}, the solution given by
$\left(\bar{R}_{\mathrm{c}}^{\ast} = 1 - \bar{\xi}_{\mathrm{c}}^{\ast},\mathbf{P}^{\ast}\right)$ satisfies the KKT optimality conditions of \eqref{Eq_Opt_ASR_RS_M}.
This can be demonstrated by following the steps in the proof of \cite[Proposition 1]{Razaviyayn2013b}, where the max-min fairness problem in a MIMO Interfering BC is considered.
Despite the different setup, the absence of RS and the assumption of perfect CSIT in \cite{Razaviyayn2013b}, the proofs are directly extendable to the problem considered here due to a key similarity: both objective functions are non-smooth and can be expressed as a max-min of sum-rate expressions.
At this point, it is concluded that solving  \eqref{Eq_Opt_AWSMSE_M} yields a solution for \eqref{Eq_Opt_ASR_RS_M}, which in turn, converges to a solution for the ASR problem in \eqref{Eq_Opt_ASR_RS} a.s. as $M \rightarrow \infty$.
\subsection{Alternating Optimization Algorithm}
\label{Section_AO}
Although problem \eqref{Eq_Opt_AWSMSE_M} is non-convex in the joint set of optimization variables, it is convex  in each of the blocks $\mathbf{P}$, $\mathbf{U}$ and $\mathbf{G}$, while fixing the other two.
Moreover, $(\mathbf{G},\mathbf{U})$ assume the closed-form MMSE solution for a given $\mathbf{P}$.
These properties are exploited using an AO algorithm.
Each iteration of the proposed algorithm consists of two steps: 1) updating $(\mathbf{G},\mathbf{U})$ for a given $\mathbf{P}$,
2) updating $\mathbf{P}$ (and $\bar{\xi}_{\mathrm{c}}$) for given $(\mathbf{G},\mathbf{U})$.
Next, each of the steps is described in detail.
\subsubsection{Updating the Equalizers and Weights}
\label{subsection_obj_fn_update}
In $n$th iteration of the AO algorithm, the equalizers and weights are updated such that
$(\mathbf{G},\mathbf{U})=\big(\mathbf{G}^{\mathrm{MMSE}}(\mathbf{P}^{[n-1]}),\mathbf{U}^{\mathrm{MMSE}}(\mathbf{P}^{[n-1]})\big)$,
where the dependency of the MMSE solution on a given precoder is highlighted in the notation, and $\mathbf{P}^{[n-1]}$ is the precoding matrix obtained in the $n-1$th iteration.
To facilitate the formulation of the precoder optimization problem  in the following step, we introduce the SAFs listed as:
$\bar{t}_{\mathrm{c},k}$, $\bar{t}_{k}$,
$\bar{\mathbf{\Psi}}_{\mathrm{c},k}$, $\bar{\mathbf{\Psi}}_{k}$,
$\bar{\mathbf{f}}_{\mathrm{c},k}$, $\bar{\mathbf{f}}_{k}$, $\bar{u}_{\mathrm{c},k}$, $\bar{u}_{k}$,
$\bar{\upsilon}_{\mathrm{c},k}$ and $\bar{\upsilon}_{k}$,
which are obtained using the updated $(\mathbf{G},\mathbf{U})$.
In particular, $\bar{u}_{\mathrm{c},k}$ and $\bar{u}_{k}$ are calculated by taking the ensemble averages over the $M$ realizations of
$u_{\mathrm{c},k}^{(m)}$ and $u_{k}^{(m)}$, respectively.
The rest are calculated in a similar manner by averaging over their corresponding realizations given by:
%
\begin{align}
\nonumber
t_{\mathrm{c},k}^{(m)}& =  u_{\mathrm{c},k}^{(m)}\left| g_{\mathrm{c},k}^{(m)}\right|^{2}
&\text{and} &&
t_{k}^{(m)} & =  u_{k}^{(m)}\left| g_{k}^{(m)}\right|^{2}\\
\nonumber
\mathbf{\Psi}_{\mathrm{c},k}^{(m)} & =  t_{\mathrm{c},k}^{(m)} \mathbf{h}_{k}^{(m)}{\mathbf{h}_{k}^{(m)}}^{H}
&\text{and} &&
\mathbf{\Psi}_{k}^{(m)} & =   t_{k}^{(m)} \mathbf{h}_{k}^{(m)}{\mathbf{h}_{k}^{(m)}}^{H}\\
\nonumber
\mathbf{f}_{\mathrm{c},k}^{(m)} & = u_{\mathrm{c},k}^{(m)} \mathbf{h}_{k}^{(m)}{g_{\mathrm{c},k}^{(m)}}^{H}
&\text{and} &&
\mathbf{f}_{k}^{(m)} & =  u_{k}^{(m)} \mathbf{h}_{k}^{(m)}{g_{k}^{(m)}}^{H} \\
\nonumber
\upsilon_{\mathrm{c},k}^{(m)} & =  \log_{2}\left(u_{\mathrm{c},k}^{(m)}\right)
&\text{and} &&
\upsilon_{k}^{(m)} & =  \log_{2}\left(u_{k}^{(m)}\right).
\end{align}
%
\subsubsection{Updating the Precoders}
\label{Subsection_Opt Pre_Vect}
Following the previous step, the problem of updating $\mathbf{P}$ is formulated by plugging $\mathbf{G}^{\mathrm{MMSE}}(\mathbf{P}^{[n-1]})$ and $\mathbf{U}^{\mathrm{MMSE}}(\mathbf{P}^{[n-1]})$ into \eqref{Eq_Opt_AWSMSE_M}.
This yields
\begin{align}
\nonumber
& \mathcal{A}_{\mathrm{RS}}^{[n]}(P_{\mathrm{t}}): \\
\label{Eq_Opt_WAMSE_QCQP}
& 
\begin{cases}
       \underset{\bar{\xi}_{\mathrm{c}}, \mathbf{P} }{\min} &
 \bar{\xi}_{\mathrm{c}} + \sum_{k=1}^{K}  \Big(
      \sum_{i=1}^{K}
      \mathbf{p}_{i}^{H}\bar{\mathbf{\Psi}}_{k}\mathbf{p}_{i}
     + \sigma_{\mathrm{n}}^{2}\bar{t}_{k}  \\
     &
     \quad \quad - 2\Re\big\{\bar{\mathbf{f}}_{k}^{H} \mathbf{p}_{k}\big\} + \bar{u}_{k}-\bar{\upsilon}_{k}
    \Big) \\
      \ \ \text{s.t.}  &
      \mathbf{p}_{\mathrm{c}}^{H}\bar{\mathbf{\Psi}}_{\mathrm{c},k}\mathbf{p}_{\mathrm{c}}
     +  \sum_{i=1}^{K}
     \mathbf{p}_{i}^{H}\bar{\mathbf{\Psi}}_{\mathrm{c},k}\mathbf{p}_{i}
     +  \sigma_{\mathrm{n}}^{2}\bar{t}_{\mathrm{c},k}
    \\
    & 
     -  2 \Re \big\{ \bar{\mathbf{f}}_{\mathrm{c},k}^{H} \mathbf{p}_{\mathrm{c}} \big\}
     +  \bar{u}_{\mathrm{c},k}
     -  \bar{\upsilon}_{\mathrm{c},k}
    \leq \bar{\xi}_{\mathrm{c}}, \ \forall k\in \mathcal{K} \\
& \|\mathbf{p}_{\mathrm{c}}\|^{2} + \sum_{k=1}^{K} \| \mathbf{p}_{k}\|^{2} \leq P_{\mathrm{t}}.
\end{cases}
\end{align}
%
The expressions of the AWMSEs in \eqref{Eq_Opt_WAMSE_QCQP} are obtained as follows.
The MSEs in \eqref{Eq_MSE} are substituted into \eqref{Eq_A_WMSEs}, followed by substituting the result into the $M$ realizations in \eqref{Eq_AWMSE_M}.
This is carried out while considering the updated equalizers and weights from the previous step, which are embedded in the SAFs listed above.
Problem \eqref{Eq_Opt_WAMSE_QCQP} is a convex Quadratically Constrained Quadratic Program (QCQP) which can be solved using  interior-point methods \cite{Boyd2004}.
\subsubsection{Algorithm and Convergence}
\label{Subsection_AO}
The two steps are repeated in an alternating manner until convergence
as summarized in Algorithm \ref{Algthm_AO}, where  $\epsilon_{R}$ is some arbitrarily small constant\footnote{Contrary to the algorithm in \cite{Razaviyayn2013a}, Algorithm \ref{Algthm_AO} applies the entire Monte-Carlo sample in each iteration. This comes at the cost of updating more equalizers and wights, yet liberates the convergence (and hence the number iterates in which $\mathbf{P}$ is updated) from the Monte-Carlo sample size. This is more suitable here as equalizers and wights are scalars obtained in closed-form, while updating $\mathbf{P}$ is the most expensive step in each iteration.}.
%
\begin{algorithm}[t]
\caption{Alternating Optimization}
\label{Algthm_AO}
\begin{algorithmic}[1]
\State \textbf{Initialize}: $n\gets 0$, $\mathcal{A}_{\mathrm{RS}}^{[n]}  \gets 0$, $\mathbf{P}$
\label{Algthm_AO_step_initialize}
\Repeat
    \State $n\gets n+1$, $\mathbf{P}^{[n-1]}\gets \mathbf{P}$
    \State $\mathbf{G}\gets \mathbf{G}^{\mathrm{MMSE}}(\mathbf{P}^{[n-1]})$, $\mathbf{U}\gets \mathbf{U}^{\mathrm{MMSE}}(\mathbf{P}^{[n-1]})$
    \label{Algthm_AO_step_MMSE}
    \State update
    $\bar{\mathbf{\Psi}}_{\mathrm{c},k},\bar{\mathbf{\Psi}}_{k},\bar{\mathbf{F}}_{\mathrm{c},k},\bar{\mathbf{F}}_{k}, \bar{t}_{\mathrm{c},k},\bar{t}_{k},\bar{u}_{\mathrm{c},k},\bar{u}_{k},\bar{\upsilon}_{\mathrm{c},k},\bar{\upsilon}_{k}$ 
    \label{Algthm_AO_step_expectations}
     \Statex\hspace{\algorithmicindent}for all $k \in \mathcal{K}$
     \State  $\mathbf{P} \gets \arg{\mathcal{A}_{\mathrm{RS}}^{[n]} }$
    \label{Algthm_AO_step_QCQP}
\Until{$\left|\mathcal{A}_{\mathrm{RS}}^{[n]} - \mathcal{A}_{\mathrm{RS}}^{[n-1]} \right| < \epsilon_{R}$ }
\end{algorithmic}
\end{algorithm}
%
\newtheorem{Proposition_WAMSE_Conv}[Proposition_Counter]{Proposition}
\begin{Proposition_WAMSE_Conv}\label{Proposition_WAMSE_Conv}
\textnormal{
For a given $\mathbb{H}^{(M)}$, the iterates generated by Algorithm \ref{Algthm_AO} converge to the set of KKT solutions of the corresponding sampled ASR problem in \eqref{Eq_Opt_ASR_RS_M}.
As $M \rightarrow \infty$, the iterates converge, a.s., to the set of KKT solutions of the ASR problem in \eqref{Eq_Opt_ASR_RS}.
}
\end{Proposition_WAMSE_Conv}
\begin{proof}
The sequence $\big\{ \mathcal{A}_{\mathrm{RS}}^{[n]} \big\}_{n=1}^{\infty}$  is monotonically decreasing, as each optimization step decreases the objective function.
Since $\mathcal{R}_{\mathrm{RS}}^{(M)}(P_{\mathrm{t}})$ is bounded above for a given power constraint, it follows that  $\mathcal{A}_{\mathrm{RS}}^{(M)}(P_{\mathrm{t}})$ is bounded below, and hence $\mathcal{A}_{\mathrm{RS}}^{[n]}$ is guaranteed to converge.
Next, we show that the corresponding sequence $\left\{\mathbf{P}^{[n]} \right\}_{n=1}^{\infty}$ converges to the set of KKT points of problem \eqref{Eq_Opt_ASR_RS_M}.
It can be seen from the relationship in \eqref{Eq_min_AWMSE_M} that for the $n$th iteration, problem $\mathcal{A}_{\mathrm{RS}}^{[n]}(P_{\mathrm{t}})$ in \eqref{Eq_Opt_WAMSE_QCQP} is in fact a convex approximation of the sampled ASR problem $\mathcal{R}_{\mathrm{RS}}^{(M)}(P_{\mathrm{t}})$ in \eqref{Eq_Opt_ASR_RS_M} around the point $\mathbf{P}^{[n-1]}$, obtained from the previous iteration.
Hence, the AO procedure in Algorithm \ref{Algthm_AO} in an instance of the Successive Convex Approximation (SCA) method in
\cite[Section 2.1]{Razaviyayn2014}.
Combining this with the conditions in \cite[Assumption 1]{Razaviyayn2014} and the fact that the iterates $\left\{\mathbf{P}^{[n]} \right\}_{n=1}^{\infty}$ lie in the compact feasible set $\mathbb{P}$, the convergence to the set of KKT points of problem \eqref{Eq_Opt_ASR_RS_M} follows from \cite[Theorem 1]{Razaviyayn2014} and \cite[Corollary 1]{Razaviyayn2013}.
The a.s. convergence to the set of KKT points of problem \eqref{Eq_Opt_ASR_RS} as $M \rightarrow \infty$ follows from  Section \ref{Subsection_SAA}.
\end{proof}
It should be highlighted that due to the non-convexity of the problem, the global optimality of the solution cannot be guaranteed in general, and different initializations of $\mathbf{P}$ may lead to different limit points. This is examined through simulations in Section \ref{subsection_conv} where it is shown that appropriate initialization yields good convergence and rate performances.
Moreover, the influence of varying the finite sample size $M$ on the performance is investigated in Section \ref{subsubsection_sample_size}.
%
\section{Conservatively Approximated WMMSE Algorithm}
\label{Section_Conservative}
In this section, we consider an alternative deterministic approximation of the AR-AWMMSE relationship in \eqref{Eq_min_AWMSE}.
This approximation is based on relaxing the dependencies of the equalizers and weights in  \eqref{Eq_min_AWMSE} on the channel state $\mathbf{H}$.
As a result, the minimizations are taken outside the expectations, and the new AWMSEs assume closed-form
expressions \cite{Negro2012,Bashar2014}.
This eliminates the SAA and the need for a possibly large Monte-Carlo sample of conditional channel realizations.
Moreover, the sets of equalizers and weights to be updated in the AO procedure reduce to two pairs for each user.
However, this simplification comes at the expense of the achievable performance, as the approximation is in fact conservative.
\subsection{Conservative Approximation}
\label{Subsection_Relaxed_AWMSEs}
The the AWMSEs are approximated by 
$\widehat{\xi}_{\mathrm{c},k} \triangleq \mathrm{E}_{\mathrm{H} \mid \widehat{\mathrm{H}} }  \big\{\xi_{\mathrm{c},k} \mid \widehat{\mathbf{H}} \big\}$ and $ \widehat{\xi}_{k}  \triangleq \mathrm{E}_{\mathrm{H} \mid \widehat{\mathrm{H}} } \big\{\xi_{k} \mid \widehat{\mathbf{H}}  \big\}$, which are expressed as
\begin{subequations}
\label{Eq_AWMSE_UB}
\begin{align}
 \widehat{\xi}_{\mathrm{c},k} &  = \widehat{u}_{\mathrm{c},k}\left(|\widehat{g}_{\mathrm{c},k}|^{2} \bar{T}_{\mathrm{c},k}  -2\Re \big\{\widehat{g}_{\mathrm{c},k}\widehat{\mathbf{h}}_{k}^{H}\mathbf{p}_{\mathrm{c}}\big\}+1\right) - \log_{2}(\widehat{u}_{\mathrm{c},k}) \\
 \widehat{\xi}_{k} & = \widehat{u}_{k}\left(|\widehat{g}_{k}|^{2}\bar{T}_{k} -2\Re \big\{\widehat{g}_{k}\widehat{\mathbf{h}}_{k}^{H}\mathbf{p}_{k}\big\}+1 \right) - \log_{2}(\widehat{u}_{k})
\end{align}
\end{subequations}
where $(\widehat{g}_{\mathrm{c},k},\widehat{g}_{k})$ and $(\widehat{u}_{\mathrm{c},k},\widehat{u}_{k})$ are the relaxed equalizers and weights respectively, and $\bar{T}_{\mathrm{c},k}$
and $\bar{T}_{k}$ are given by
\begin{subequations}
\label{Eq_T_bar}
\begin{align}
\label{Eq_T_c_bar}
\bar{T}_{\mathrm{c},k} & =  \mathbf{p}_{\mathrm{c}}^{H}\big(\widehat{\mathbf{h}}_{k}\widehat{\mathbf{h}}_{k}^{H} + \mathbf{R}_{\mathrm{e},k}\big)\mathbf{p}_{\mathrm{c}} + \bar{T}_{k} \\
\label{Eq_T_k_bar}
\bar{T}_{k} &  =  \sum_{i=1}^{K}\mathbf{p}_{i}^{H}\big(\widehat{\mathbf{h}}_{k}\widehat{\mathbf{h}}_{k}^{H} + \mathbf{R}_{\mathrm{e},k}\big)\mathbf{p}_{i} + \sigma_{\mathrm{n}}^{2}.
\end{align}
\end{subequations}
The optimum relaxed equalizers are derived from \eqref{Eq_AWMSE_UB} as:
$\widehat{g}_{\mathrm{c},k}^{\mathrm{MMSE}} = \mathbf{p}_{\mathrm{c}}^{H}\widehat{\mathbf{h}}_{k}\bar{T}_{\mathrm{c},k}^{-1}$
and
$\widehat{g}_{k}^{\mathrm{MMSE}} = \mathbf{p}_{k}^{H}\widehat{\mathbf{h}}_{k}\bar{T}_{k}^{-1}$.
Plugging them back into \eqref{Eq_AWMSE_UB}, the optimum relaxed weights are obtained as
$\widehat{u}_{\mathrm{c},k}^{\mathrm{MMSE}} = \big( \widehat{\varepsilon}_{\mathrm{c},k}^{\mathrm{MMSE}} \big)^{-1}$
and
$\widehat{u}_{k}^{\mathrm{MMSE}} = \big( \widehat{\varepsilon}_{k}^{\mathrm{MMSE}} \big)^{-1}$, where
$\widehat{\varepsilon}_{\mathrm{c},k}^{\mathrm{MMSE}} \triangleq 1 - \bar{T}_{\mathrm{c},k}^{-1} |\widehat{\mathbf{h}}_{k}^{H}\mathbf{p}_{\mathrm{c}}|^{2}  $
and
$\widehat{\varepsilon}_{k}^{\mathrm{MMSE}} \triangleq 1 - \bar{T}_{k}^{-1} |\widehat{\mathbf{h}}_{k}^{H}\mathbf{p}_{k}|^{2} $.
It is clear that the relaxed equalizers and weights are functions of $\widehat{\mathbf{H}}$.
Incorporating this approximation into the AR-AWMMSE relationship in \eqref{Eq_min_AWMSE}, the expectations are taken inside the minimizations such that
\begin{subequations}
\label{Eq_min_relaxed_AWMSE}
\begin{align}
\widehat{\xi}_{\mathrm{c},k}^{\mathrm{MMSE}} & \triangleq \underset{\widehat{u}_{\mathrm{c},k}, \widehat{g}_{\mathrm{c},k}}{\min}
\mathrm{E}_{\mathrm{H} \mid \widehat{\mathrm{H}} }  \big\{\xi_{\mathrm{c},k} \mid \widehat{\mathbf{H}} \big\}
 = 1 - \widehat{R}_{\mathrm{c},k}
\\
\widehat{\xi}_{k}^{\mathrm{MMSE}} & \triangleq \underset{\widehat{u}_{k}, \widehat{g}_{k}}{\min} \;
\mathrm{E}_{\mathrm{H} \mid \widehat{\mathrm{H}} }  \big\{\xi_{k} \mid \widehat{\mathbf{H}} \big\} = 1 - \widehat{R}_{k}
\end{align}
\end{subequations}
where  $\widehat{R}_{\mathrm{c},k} \triangleq -\log_{2}(\widehat{\varepsilon}_{\mathrm{c},k}^{\mathrm{MMSE}})$ and
$\widehat{R}_{k} \triangleq -\log_{2}(\widehat{\varepsilon}_{k}^{\mathrm{MMSE}})$.
The approximations in \eqref{Eq_min_relaxed_AWMSE} are upper-bounds for the expressions in \eqref{Eq_min_AWMSE}, i.e.
$\widehat{\xi}_{\mathrm{c},k}^{\mathrm{MMSE}} \geq \bar{\xi}_{\mathrm{c},k}^{\mathrm{MMSE}}$
and
$\widehat{\xi}_{k}^{\mathrm{MMSE}}  \geq \bar{\xi}_{k}^{\mathrm{MMSE}}$.
This follows from the fact that moving the expectation inside the minimization does not decrease the value.
Hence, we have
\begin{equation}
\label{Eq_AR_LB}
\widehat{R}_{\mathrm{c},k} \leq \bar{R}_{\mathrm{c},k}
 \quad \text{and} \quad
\widehat{R}_{k}  \leq \bar{R}_{k}.
\end{equation}
By employing \eqref{Eq_AWMSE_UB}, a conservative version of the AWSMSE problem is formulated as
\begin{equation}
 \label{Eq_Opt_AWSMSE_UB}
\widehat{\mathcal{A}}_{\mathrm{RS}}(P_{\mathrm{t}}):
\begin{cases}
       \underset{\widehat{\xi}_{\mathrm{c}}, \mathbf{P}, \widehat{\mathbf{u}}, \widehat{\mathbf{g}} }{\min} &
 \widehat{\xi}_{\mathrm{c}}+ \sum_{k=1}^{K} \widehat{\xi}_{k}  \\
      \ \ \text{s.t.}  & \widehat{\xi}_{\mathrm{c},k} \leq \widehat{\xi}_{\mathrm{c}}, \; \forall k\in\mathcal{K} \\
                    & \mathrm{tr}\big(\mathbf{P}\mathbf{P}^{H}\big) \leq P_{\mathrm{t}}
\end{cases}
\end{equation}
%
where $\widehat{\mathbf{u}} \triangleq \big\{ \widehat{u}_{\mathrm{c},k},\widehat{u}_{k} \mid k \in \mathcal{K} \big\}$
and
$\widehat{\mathbf{g}} \triangleq \big\{ \widehat{g}_{\mathrm{c},k},\widehat{g}_{k} \mid k \in \mathcal{K} \big\}$.
Using the same argument in Section \ref{Subsection_Augmented_AWSMSE}, it can be shown that the aforementioned MMSE solution is optimal for \eqref{Eq_Opt_AWSMSE_UB}, and the cost function at any optimum point is an affine transformation of the conservative ASR given by
$\min_{j} \; \big\{\widehat{R}_{\mathrm{c},j} \big\}_{j=1}^{K} + \sum_{k=1}^{K} \widehat{R}_{k}$.
Problem \eqref{Eq_Opt_AWSMSE_UB} is solved by slightly modifying Algorithm \ref{Algthm_AO}. Specifically, $(\mathbf{G},\mathbf{U})$ are reduced to $(\widehat{\mathbf{g}},\widehat{\mathbf{u}})$, and step \ref{Algthm_AO_step_expectations} is eliminated as \eqref{Eq_AWMSE_UB} is directly employed to formulate the QCQP.
\subsection{Conservative Performance Limitations}
\label{Subsection_conservative_limitations}
It can be seen that  \eqref{Eq_Opt_AWSMSE_UB} is a restricted version of the AWSMSE problem in \eqref{Eq_Opt_AWSMSE_M}, obtained
by restricting the domain of \eqref{Eq_Opt_AWSMSE_M} such that the sets of equalizers and weights remain unchanged across the $M$ realizations, and driving $M\rightarrow \infty$.
By definition, solving \eqref{Eq_Opt_AWSMSE_UB} is equivalent to maximizing a lower-bound on the ASR.
Hence, the conservative ESR given by $K + 1 - \mathrm{E}_{\widehat{\mathrm{H}}} \big\{\widehat{\mathcal{A}}_{\mathrm{RS}}(P_{\mathrm{t}}) \big\}$ is achievable.
As the equalizers are updated using partial CSIT, the availability of highly accurate CSIR is completely ignored by the BS in the design (also called an ignorant approach \cite{Joudeh2014}), which may lead to a loss in performance compared to the SAA method.
This leaves us wondering about the effectiveness of the conservative approximation, i.e. the looseness of the lower-bound.
A similar ignorant approach was used for worst-case robust designs under bounded errors, where it was observed that the rate lower-bound fails to achieve the anticipated DoF \cite{Tajer2011}.
This was analysed in \cite{Joudeh2016a}, where it was shown that the loss in performance is due to self-interference terms induced by the approximation.
The same argument applies here, where it can be seen that the equivalent conservative SINRs, defined as
$\widehat{\gamma}_{\mathrm{c},k} = \big(1-\widehat{\varepsilon}_{\mathrm{c},k}^{\mathrm{MMSE}} \big)/\widehat{\varepsilon}_{\mathrm{c},k}^{\mathrm{MMSE}}$
and
$\widehat{\gamma}_{k} = \big(1-\widehat{\varepsilon}_{k}^{\mathrm{MMSE}} \big)/\widehat{\varepsilon}_{k}^{\mathrm{MMSE}}$,
have the self-interference terms, given by $\mathbf{p}_{\mathrm{c}}^{H} \mathbf{R}_{\mathrm{e},k}\mathbf{p}_{\mathrm{c}}$ and
$\mathbf{p}_{k}^{H} \mathbf{R}_{\mathrm{e},k}\mathbf{p}_{k}$ respectively, within their denominators.
Such terms may scale significantly with increased power, depending on the CSIT quality, hence limiting the rate performance as we observe in Section \ref{Subsubsection_conservative_simulations}.
It is worth noting that while relaxing the dependencies of equalizers on $\mathbf{H}$ is carried out by the BS to simplify the design algorithm, is not necessarily followed by receivers which are still expected to apply perfect CSIR to update their equalizers.
However, predicting the possibly higher ERs (achievable by the users) at the BS, and hence adjusting the transmission rates, requires the evaluation of the expressions in \eqref{Eq_ER_EAR}  and \eqref{Eq_ERc_EARc}, which in turn requires numerical calculations that bring us back to Monte-Carlo samplings. Hence for this approach, we assume that the BS transmit at the conservative ERs.
\section{Numerical Results and Analysis}
\label{Section_Numerical_Results}
In this section, the proposed algorithms are evaluated through simulations.
The channel $\mathbf{H}$ has i.i.d. complex Gaussian entries with unit variance, i.e. $\mathcal{C}\mathcal{N}\left(0,1\right)$.
The noise variance is fixed as $\sigma_{\mathrm{n}}^{2} = 1$, from which the long-term SNR is given as $P_{\mathrm{t}}$.
Entries of $\widetilde{\mathbf{H}}$ are also i.i.d complex Gaussian drawn from $\mathcal{C}\mathcal{N}\left(0,\sigma_{\mathrm{e}}^{2}\right)$, where
$ \sigma_{\mathrm{e}}^{2} = N_{\mathrm{t}}^{-1}\sigma_{\mathrm{e},k}^{2}, \; \forall k \in \mathcal{K}$.
The error variance is given as
$\sigma_{\mathrm{e}}^{2} = \beta P_{\mathrm{t}}^{-\alpha}$, where $\beta \geq 0$ and $\alpha \in [0,1]$ are varied to represent different CSIT accuracies and SNR scalings.
It follows that the channel estimate $\widehat{\mathbf{H}} = \mathbf{H} - \widetilde{\mathbf{H}}$ is also Gaussian.
The channel realization $\mathbf{H}$ should not be confused with a conditional realization $\mathbf{H}^{(m)}$.
While the former represents the actual channel experienced by the users and unknown to the BS, the latter is part of a sample $\mathbb{H}^{(M)}$ available at the BS and used to calculate the SAFs.
The size of the sample is set to $M=1000$ throughout the simulations, unless otherwise stated.
For a given estimate $\widehat{\mathbf{H}}$, the $m$th conditional realization is obtained as $\mathbf{H}^{(m)}  = \widehat{\mathbf{H}} +  \widetilde{\mathbf{H}}^{(m)}$, where $\widetilde{\mathbf{H}}^{(m)}$ is drawn from the error distribution.
More on generating the channel realizations in the simulations is given in Appendix \ref{Appendix_Channel}.
Convex optimization problems are solved using the CVX toolbox \cite{Grant2008}.
\subsection{Convergence}
\label{subsection_conv}
\begin{figure}[t]
\centering
\includegraphics[width = 0.45\textwidth]{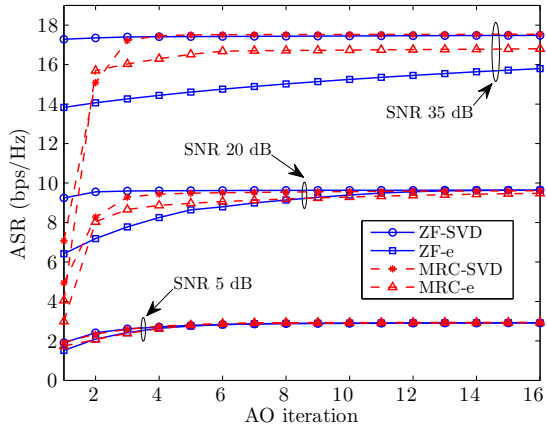}
\caption{ASR convergence of Algorithm \ref{Algthm_AO} using four different initializations of $\mathbf{P}$ for  1 randomly generated channel estimate, $K,N_{\mathrm{t}}=2$, $\sigma_{\mathrm{e}}^{2} = P_{\mathrm{t}}^{-0.6}$, and $\mathrm{SNR} = 5$, $20$ and $35$ dB.}
\label{Fig_Convergence}
\end{figure}
First, we examine the convergence of Algorithm \ref{Algthm_AO} under four different initializations of $\mathbf{P}$.
The first initialization, denoted by ZF-e, is taken as the DoF-motivated design in
\cite{Yang2013,Hao2013}. To recall, we have $\mathbf{p}_{\mathrm{c}} = \sqrt{q_{\mathrm{c}}}\widehat{\mathbf{p}}_{\mathrm{c}}$
and $\mathbf{p}_{k} = \sqrt{q_{k}}\widehat{\mathbf{p}}_{k}$, where $q_{\mathrm{c}} = P_{\mathrm{t}} - P_{\mathrm{t}}^{\alpha}$ and
$q_{k} = P_{\mathrm{t}}^{\alpha}/K$, $\widehat{\mathbf{p}}_{\mathrm{c}} = \mathbf{e}_{1}$
and  $[\widehat{\mathbf{p}}_{1},\ldots,\widehat{\mathbf{p}}_{K}]$ are normalized ZF-BF vectors constructed using $\widehat{\mathbf{H}}$.
The second initialization, labeled as ZF-SVD, is a modification of ZF-e, where $\widehat{\mathbf{p}}_{c}$ is slightly optimized by choosing it as the dominant left singular vector of $\widehat{\mathbf{H}}$.
The third and fourth initializations, labeled as MRC-e and MRC-SVD respectively, maintain the same common precoders as ZF-e and ZF-SVD respectively. However, Maximum Ratio Combining (MRC) is employed instead of ZF-BF, i.e. $\widehat{\mathbf{p}}_{k} = \widehat{\mathbf{h}}_{k} \backslash \| \widehat{\mathbf{h}}_{k} \|$.
The ASR convergence of Algorithm \ref{Algthm_AO} for $\sigma_{\mathrm{e}}^{2} = P_{\mathrm{t}}^{-0.6}$ and SNRs $5$, $20$ and $35$ dB is shown in Fig. \ref{Fig_Convergence}.
It is evident that the algorithm converges to a limit point regardless of the initialization.
However, the speed of convergence is influenced by the initial state, which may also influence the limit point as the problem is non-convex.
The initialization effect becomes more visible as SNR grows large.
For example, initializing the common precoder using SVD enhances the convergence at high SNRs.
In the following results, MRC-SVD is adopted as it provides good overall performance over various channel realizations and a wide range of SNRs.
%
\subsection{Ergodic Sum Rate Performance}
\label{subsection_results_SR}
\begin{figure}[t!]
\centering
\includegraphics[width = 0.45\textwidth]{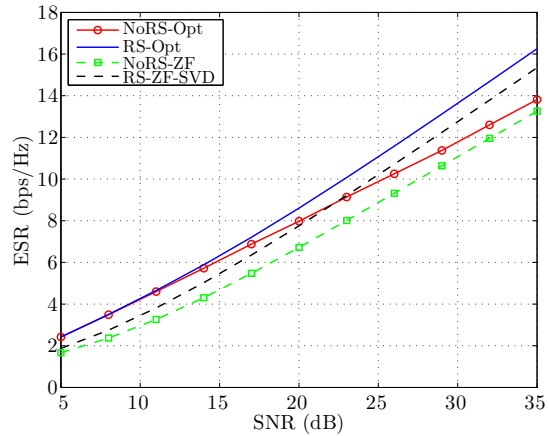}
\caption{Comparison between the ESR performances of NoRS and RS transmission schemes for $K,N_{\mathrm{t}}=2$ and $\sigma_{\mathrm{e}}^{2}=P_{\mathrm{t}}^{-0.6}$.}
\label{Fig_ESR_06}
\end{figure}
\begin{figure}[t!]
\centering
\includegraphics[width = 0.45\textwidth]{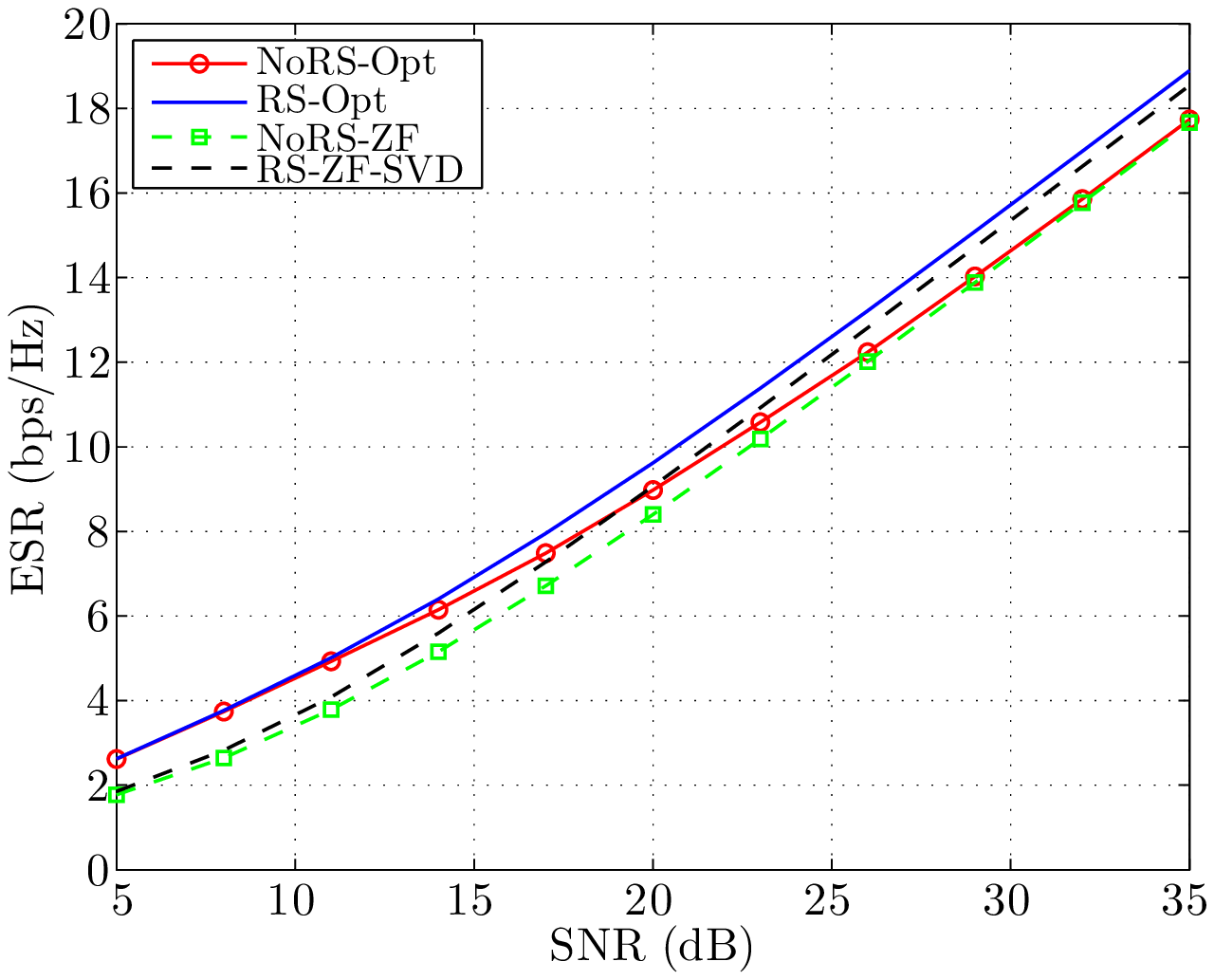}
\caption{Comparison between the ESR performances of NoRS and RS transmission schemes for $K,N_{\mathrm{t}}=2$ and $\sigma_{\mathrm{e}}^{2}=P_{\mathrm{t}}^{-0.9}$. }
\label{Fig_ESR_09}
\end{figure}
\begin{figure}[t!]
\centering
\includegraphics[width = 0.45\textwidth]{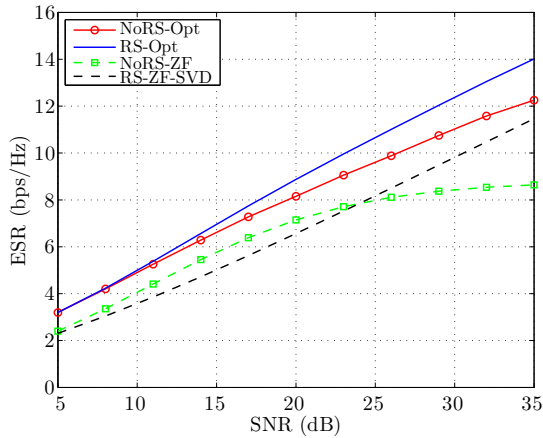}
\caption{Comparison between the ESR performances of NoRS and RS transmission schemes for $K,N_{\mathrm{t}}=2$ and $\sigma_{\mathrm{e}}^{2} = 0.063$. }
\label{Fig_ESR_Sigma_063}
\end{figure}
%
Here we evaluate the achievable ESRs obtained by averaging the ASRs over $100$ channel realizations.
For each channel realization, Algorithm \ref{Algthm_AO} is used to obtain the RS ASR, and a modification of the algorithm where the common precoder is switched off is used for NoRS.
\subsubsection{RS vs. NoRS}
The performance of the optimized RS scheme (RS-Opt) is compared to the optimized NoRS scheme (NoRS-Opt).
ZF-BF with Water-Filling (WF), where power allocation is carried out assuming that the estimate $\widehat{\mathbf{H}}$ is perfect, is considered as a baseline for NoRS transmission. This scheme is termed as NoRS-ZF.
On the other hand, we consider a modified version of the ZF-SVD initialization in the previous subsection as a baseline for RS transmission.
In particular, the power splitting between the common message and the private messages is maintained, while WF is used to allocate the power among the private messages. This scheme is termed RS-ZF-SVD.
The 2-user ESRs obtained using the different optimization schemes for $\sigma_{\mathrm{e}}^{2}=P_{\mathrm{t}}^{-0.6}$ and $P_{\mathrm{t}}^{-0.9}$ are shown in Fig. \ref{Fig_ESR_06} and Fig. \ref{Fig_ESR_09}, respectively.
Comparing RS-Opt and NoRS-Opt, the two schemes perform similarly at low SNRs. In particular, both reduce to SU transmission by switching off the weaker user.
As SNR grows, MU transmission starts to take over.
However, reducing MU interference to the level of noise (or eliminating it) is not possible due to imperfect CSIT.
Therefore, the overall performance does not necessarily benefit from additional power inaccurately directed towards a given user, as it may cause more damage by interfering with other users.
At this stage, RS-Opt starts to part from NoRS-Opt by switching on the common message.
The contribution of the common message primarily manifests at high SNRs, where the gap between RS-Opt and NoRS-Opt  exceeds $4$ dB for $\alpha = 0.6$.
For $\alpha = 0.9$, which is almost perfect from a DoF perspective, RS is not as instrumental as it is for $\alpha = 0.6$. However, rate gains can still be observed at high SNRs.
Regarding the baseline schemes, their inferiority compared to optimized schemes is evident over the entire SNR range.
In the low SNR regime, baselines fall behind due to the strict application of ZF-BF, which is not ideal in this case.
However, the RS-ZF-SVD scheme slightly compensates for this through the common message.
However, it fails to match the optimized transmission in RS-Opt and NoRS-Opt.
In the high SNR regime, each  baseline scheme achieves the same DoF (slope of the curve) as its optimized counterpart with a rate gap, which is particularly noticeable for $\alpha = 0.6$.

The performance under CSIT errors that do not scale with SNR ($\alpha = 0$) is shown in Fig. \ref{Fig_ESR_Sigma_063}, where $
\sigma_{\mathrm{e}}^{2} = \big( 10^{\frac{20}{10}} \big)^{-0.6} = 0.063$, which corresponds to the CSIT quality obtained at $20$ dB SNR  w.r.t Fig. \ref{Fig_ESR_06}.
First, it can be seen that the rate of NoRS-ZF saturates. This is due to the naive employment of $\widehat{\mathbf{H}}$ to design the ZF-BF vectors and allocate power as if it was a perfect estimate. At high SNRs, residual interference dominates and caps the performance.
RS-ZF-SVD employs the DoF-motivated power splitting (i.e. $q_{\mathrm{c}} = P_{\mathrm{t}}$ and $q_{k} = 0$) and hence reduces to multicast transmission (sending only the common message) maintaining a DoF of 1.
On the other hand, NoRS-Opt and RS-Opt schemes achieve significantly better performances compared to their corresponding baselines, over the entire  range of SNRs.
Moreover, although  RS-Opt is not expected to achieve DoF gains over NoRS-Opt as $P_{\mathrm{t}} \rightarrow \infty$, the former still manages to deliver superior rate performance at medium and high SNRs.
%
\subsubsection{Increased Number of Users}
%
\begin{figure}
\centering
\includegraphics[width = 0.45\textwidth]{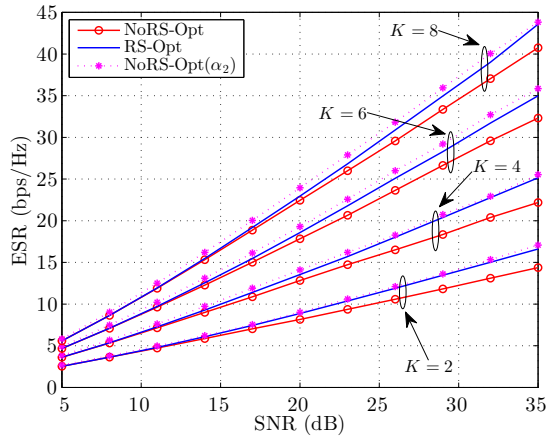}
\caption{ESR performances of NoRS-Opt and RS-Opt with $\sigma_{\mathrm{e}}^{2} = P_{\mathrm{t}}^{-\alpha}$,
  and NoRS-Opt with $\sigma_{\mathrm{e}}^{2} =  P_{\mathrm{t}}^{-\alpha_{2}}$, where $\alpha_{2}  =  \frac{1 + (K-1)\alpha }{K}$, for $\alpha = 0.6$, and $K,N_{\mathrm{t}}=2$, $4$,$6$ and $8$.}
\label{Fig_ESR_06_K4_K6}
\end{figure}
Theorem \ref{Theorem_RS_DoF} implies that the relative DoF gain achieved by RS over NoRS decreases as $K$ increases.
Fig. \ref{Fig_ESR_06_K4_K6} shows the ESRs achieved by NoRS-Opt and RS-Opt for $K=2$, $4$, $6$ and $8$  users, assuming $\alpha = 0.6$.
Fig. \ref{Fig_ESR_06_K4_K6} also shows the NoRS-Opt performance for a CSIT error exponent given as: $\alpha_{2} = \frac{1 + (K-1)\alpha }{K}$.
This corresponds to a conventional system that achieves the same sum DoF as the RS system, but at the cost of higher CSIT quality requirements since $\alpha_{2} > \alpha, \; \forall \alpha \in (0,1)$. It is evident that the two perform closely, which highlights another benefit of RS, i.e. relaxed CSIT requirements compared to NoRS.
%
\subsubsection{SAA Sample Size}
\label{subsubsection_sample_size}
%
\begin{figure}[t!]
\centering
\includegraphics[width = 0.45\textwidth]{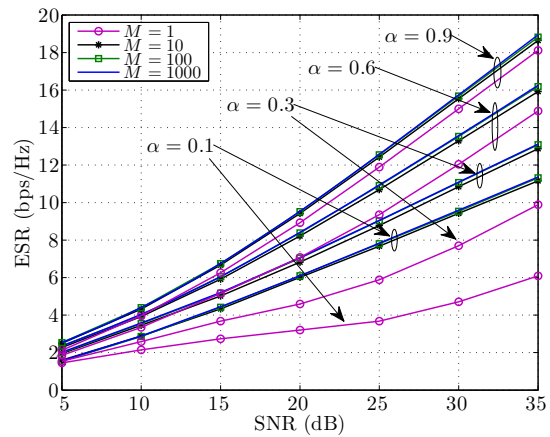}
\caption{The effect of changing $M$ on the ESR performance of the RS SAA algorithm for $K,N_{\mathrm{t}} = 2$, $\sigma_{\mathrm{e}}^{2} =  P_{\mathrm{t}}^{-\alpha}$, and $\alpha = 0.1$, $0.3$, $0.6$ and $0.9$.}
\label{Fig_ESR_M}
\end{figure}
Next, we look at the influence of changing the sample size $M$ on the SAA algorithm.
The 2-user ESRs achieved by solving Algorithm \ref{Algthm_AO} for $M = 1,10,100$ and $1000$ are given in Fig. \ref{Fig_ESR_M}.
Different scaling CSIT qualities are considered where $\sigma_{\mathrm{e}}^{2} = P_{\mathrm{t}}^{-\alpha}$, and $\alpha = 0.1,0.3,0.6$ and $0.9$.
Using a sample of one realization significantly degrades the performance, as the resulting design lacks statistical knowledge of the CSIT uncertainty.
On the other hand, the SAA algorithm performs very well with a sample size as small as $M=10$ under the specified settings. Higher sample sizes of $M=100$ and $1000$ perform almost identically.
Note that the achievable ESRs in this part are obtained by numerically evaluating the right-most expressions in \eqref{Eq_ER_EAR} and \eqref{Eq_ERc_EARc}, as taking the expectations of the sampled ASR objective functions for low $M$s is not reflective of the achievable performance.
\subsubsection{Conservative Algorithm}
\label{Subsubsection_conservative_simulations}
%
\begin{figure}
\centering
\includegraphics[width = 0.45\textwidth]{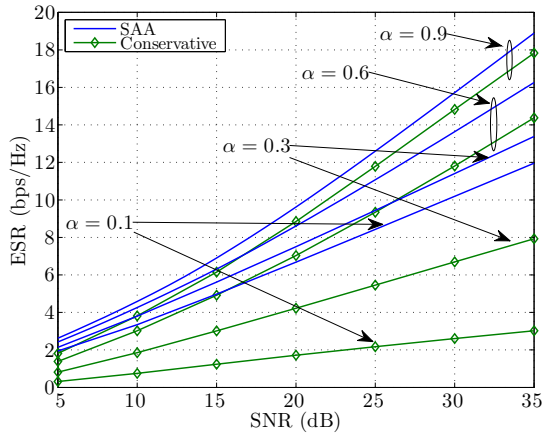}
\caption{ESR performances of the SAA and conservative approaches for $K,N_{\mathrm{t}} = 2$, $\sigma_{\mathrm{e}}^{2} =  P_{\mathrm{t}}^{-\alpha}$, and $\alpha = 0.1$, $0.3$, $0.6$ and $0.9$.}
\label{Fig_ESR_LB}
\end{figure}
In this part, we look at the performance of the conservative design in Section \ref{Section_Conservative}.
For the Gaussian CSIT error described earlier in this section, we have $\mathbf{R}_{\mathrm{e},k} = \sigma_{\mathrm{e}}^{2} \mathbf{I}, \; \forall k \in \mathcal{K}$.
The conservative ESR is obtained by averaging the conservative ASRs obtained by solving \eqref{Eq_Opt_AWSMSE_UB}.
The ESRs predicted by the BS using the SAA approach and the conservative approach are given in Fig. \ref{Fig_ESR_LB} for $K=2$ and $\sigma_{\mathrm{e}}^{2} = P_{\mathrm{t}}^{-\alpha}$, with $\alpha = 0.1,0.3,0.6$ and $0.9$.
The SAA scheme achieves significant gains over the conservative scheme, which increase with decreased CSIT qualities.
This is explained by the discussion in Section \ref{Subsection_conservative_limitations}.
In particular, self-interference terms become dominant and the ASR lower-bound becomes looser.
This effect is extremely detrimental for low CSIT qualities, as the achievable ESRs are highly undermined by the BS.
%
\subsection{Ergodic Rate Region}
\label{Subsection_Rate_Region}
\begin{figure}[t]
\centering
\includegraphics[width = 0.42\textwidth]{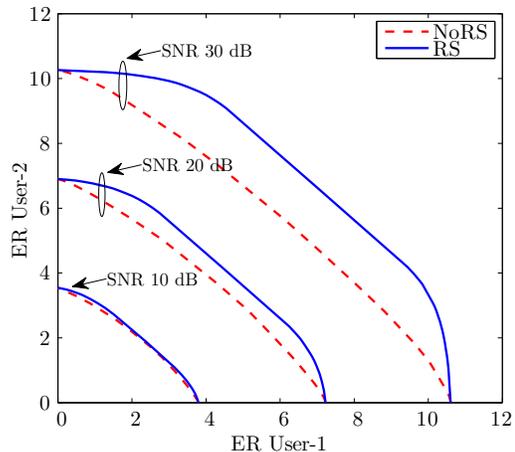}
\caption{2-user ER regions for $\sigma_{\mathrm{e}}^{2}=P_{\mathrm{t}}^{-0.6}$, and $\mathrm{SNR}= 10$, $20$ and $30$ dB.}
\label{Fig_Region}
\end{figure}
In the final part, we investigate the NoRS and RS achievable ER regions in a two-user scenario.
ER regions are obtained by solving Weighted ESR (WESR) problems, where different boundary points are realized by varying the weights.
For NoRS, this is achieved by incorporating the weights into the ASR problem in \eqref{Eq_Opt_ASR_NoRS}.
For a fixed pair of weights, the problem is solved for several channel realizations. A boundary ER region point is obtained by averaging the resulting AR realizations for a given pair of weights.
For the RS problem, applying the weights directly to \eqref{Eq_Opt_ASR_RS} results in a region with the common message allocated to one user the entire time, e.g. $k_{\mathrm{RS}} = 1$.
A second region is obtained by allocating the common message to the other user, and the full region is given by the convex-hull enclosing the two regions by time-sharing of the extremity points.
Alternatively, all boundary points for the RS region can be obtained by assuming that $\bar{R}_{\mathrm{c}}$ is shared between users such that $\bar{C}_{k}$ is the $k$th user's portion of the common AR with $\sum_{k=1}^{K}\bar{C}_{k} = \bar{R}_{\mathrm{c}}$.
The WASR problem is formulated as
\begin{equation}
\label{Eq_Opt_WASR_RS}
\mathcal{R}_{\mathrm{WRS}}(P_{\mathrm{t}}):
\begin{cases}
       \underset{\mathbf{P},\{\bar{C}_{k}\}_{k=1}^{K}}{\max} &
       \sum_{k=1}^{K} w_{k}(\bar{R}_{k}+\bar{C}_{k})  \\
       \text{s.t.}  & \bar{R}_{\mathrm{c},k} \geq \sum_{j=1}^{K} \bar{C}_{j} , \; \forall k\in\mathcal{K} \\
                    & \bar{C}_{k} \geq 0, \ \forall k \in \mathcal{K} \\
                    & \mathrm{tr}\big(\mathbf{P}\mathbf{P}^{H}\big) \leq P_{\mathrm{t}}
\end{cases}
\end{equation}
where the weights $\{w_{k} \mid k \in \mathcal{K} \}$ are fixed parameters.
\eqref{Eq_Opt_WASR_RS} is solved by formulating its equivalent AWSMSE problem, and modifying Algorithm \ref{Algthm_AO} accordingly.
To obtain the two-user regions shown in Fig. \ref{Fig_Region}, the corresponding problems are solved for $43$ different pairs of weights.
The first weight is fixed as $w_{1} = 1$, and the second weight changes as
$w_{2} \in \{10^{-3},10^{-1},10^{-0.95},\ldots,10^{0.95},10,10^{3}\}$.
What is meaningful for each pair is the ratio between the weights rather than their absolute levels \cite{Christensen2008}.
Each point on the ER region is characterized by the tuple
$ \mathrm{E}_{\widehat{\mathrm{H}}}\big\{(\bar{R}_{1} ,\bar{R}_{2})\big\}$ for NoRS, and
$ \mathrm{E}_{\widehat{\mathrm{H}}}\big\{(\bar{R}_{1} + \bar{C}_{1} ,\bar{R}_{2} + \bar{C}_{2}  )\big\}$ for RS, where each pair of ARs is obtained from solving the corresponding WASR problem\footnote{More accurately, we take the convex hulls enclosing the resulting points.}.

The ER regions shown in Fig. \ref{Fig_Region} are obtained for SNRs: $10$, $20$ and $30$ dB, and $\sigma_{\mathrm{e}}^{2} = P_{\mathrm{t}}^{-0.6}$.
As expected, the gap between RS and NoRS grows with SNR.
The performance for $30$ dB SNR is particularly interesting, as it is evident that RS enlarges the whole ER region significantly.
For example, while guaranteeing an ER of $10$ bps/Hz for user-1, an ER of almost $4$ bps/Hz can be achieved by user-2.
On the other hand, guaranteeing the same user-1 ER using NoRS restricts the ER of user-2 to just over $1.5$ bps/Hz.
This observation is of special interest for designs exploiting different points of the rate regions, e.g. max-min fairness and Quality of Service (QoS) based designs as seen in \cite{Joudeh2015a,Joudeh2016a}.
\section{Conclusion}
\label{Section_conclusion}
The problem of ESR maximization in MU-MISO systems with linear precoding and partial CSIT was addressed by employing the RS transmission strategy, where part of the MU interference is broadcasted to (and decoded by) all users.
The precoding matrix is optimized based on the available channel estimate to maximized a conditional ASR metric, computed using partial CSIT knowledge.
The stochastic ASR problem was transformed into a deterministic counterpart using the SAA method. This was followed by applying the Rate-WMMSE transformation and an AO algorithm, which converges to a stationary solution.
An alternative deterministic approximation of the ASR problem was developed based on the conservative method in \cite{Negro2012,Bashar2014}.
While this approach avoids the sampling process, its guaranteed achievable performance is shown to be inferior.
The effectiveness of the proposed algorithms and the benefits of adopting the RS strategy were demonstrated through simulations.
In particular, it was shown that the gains anticipated by the DoF analysis can be realized at finitely high SNRs where RS designs outperform their NoRS counterparts.
Moreover, additional benefits of RS including relaxed CSIT quality requirements were highlighted.
Finally, the two-user ER region was numerically obtained by solving a sequence of WASR problems. RS proved  to enlarge the entire achievable ER region, drawing attention to the potentials of employing RS in QoS constrained designs.
\appendices
\section{Proof of Theorem \ref{Theorem_RS_DoF}}
\label{Appendix_Proof_Theorem_RS_DoF}
Since $\mathbf{P}$ is separately designed for each $\widehat{\mathbf{H}}$, we focus on a precoding scheme defined for a given $\widehat{\mathbf{H}}$ as $\{\mathbf{P}\}_{P_{\mathrm{t}}}$.
The associated powers allocated to the common precoder and the $k$th private precoder are given as:
$q_{\mathrm{c}} \triangleq \| \mathbf{p}_{\mathrm{c}} \|^{2}$ and $q_{k} \triangleq \| \mathbf{p}_{k} \|^{2}$ respectively,
which scale with increased $P_{\mathrm{t}}$ as $O(P_{\mathrm{t}}^{a_{\mathrm{c}}})$ and  $O(P_{\mathrm{t}}^{a_{k}})$ respectively, where $a_{\mathrm{c}},a_{k} \in [0,1]$\footnote{Note that $a_{\mathrm{c}},a_{k} < 0$ can be replaced with $a_{\mathrm{c}},a_{k} =0$ without influencing the DoF results derived next.}.
The streams interfering with the $k$th user are dominated by a power scaling factor of $\bar{a}_{k} \triangleq \max_{j}\{a_{j}\}_{j \neq k} $.
We define the corresponding $k$th user's conditional average DoFs as
\begin{equation}
\label{Eq_DoF}
\bar{d}_{\mathrm{c},k}   \triangleq     \lim_{P_{\mathrm{t}}\rightarrow\infty} \frac{
 \bar{R}_{\mathrm{c},k}(P_{\mathrm{t}})  }{\log_{2}(P_{\mathrm{t}})}
\quad \text{and} \quad
\bar{d}_{k}  \triangleq  \lim_{P_{\mathrm{t}}\rightarrow\infty} \frac{
 \bar{R}_{k}(P_{\mathrm{t}})  }{\log_{2}(P_{\mathrm{t}})}.
\end{equation}
For a given precoding scheme defined over all $\widehat{\mathbf{H}}$, the $k$th user's long-term achievable DoFs are given by $d_{\mathrm{c},k}   \triangleq    \lim_{P_{\mathrm{t}}\rightarrow\infty} \frac{
\mathrm{E}_{\widehat{\mathrm{H}}} \big\{ \bar{R}_{\mathrm{c},k}(P_{\mathrm{t}}) \big\} }{\log_{2}(P_{\mathrm{t}})} =
\mathrm{E}_{\widehat{\mathrm{H}}} \{ \bar{d}_{\mathrm{c},k} \}$
and
$d_{k}  \triangleq  \lim_{P_{\mathrm{t}}\rightarrow\infty} \frac{
\mathrm{E}_{\widehat{\mathrm{H}}} \big\{ \bar{R}_{k}(P_{\mathrm{t}}) \big\} }{\log_{2}(P_{\mathrm{t}})}
= \mathrm{E}_{\widehat{\mathrm{H}}} \{ \bar{d}_{k} \}$,
which follows from the bounded convergence theorem.
The same definitions extend to NoRS precoding schemes while discarding the common part.
Without loss of generality, $\sigma_{\mathrm{n}}^{2} = 1$ is assumed throughout the proof.
Each of the results in \eqref{Eq_DoF_no_RS} and \eqref{Eq_DoF_RS} is obtained through two steps.
First, we show that the DoF is upper-bounded by its corresponding result. Then, we show that the upper-bounds are achievable via feasible precoding schemes.
Before proceeding, we define the function $(x)^{+} \triangleq \max\{x,0\}$.
\subsubsection{Proof of \eqref{Eq_DoF_no_RS}}
For an arbitrary precoding scheme,
let us first find an upper-bound for
\begin{equation}
\label{Eq_AR_k_DoF_proof}
\bar{R}_{k} =
\mathrm{E}_{\mathrm{H} \mid \widehat{\mathrm{H}}} \big\{ \log_{2}(T_{k}) - \log_{2}(I_{k}) \mid \widehat{\mathbf{H}} \big\}
\end{equation}
by upper-bounding and lower-bounding the first and second right-hand-side terms respectively.
From Jensen's inequality, we write
$\mathrm{E}_{\mathrm{H} \mid \widehat{\mathrm{H}}} \big\{ \log_{2}(T_{k}) \mid \widehat{\mathbf{H}} \big\}  \leq \log_{2}( \bar{T}_{k})$,
where $\bar{T}_{k}$ is defined in \eqref{Eq_T_k_bar}.
From the Cauchy-Schwarz inequality and the isotropic property of the CSIT errors, we have
$\bar{T}_{k} \leq  \big( \|\widehat{\mathbf{h}}_{k} \|^{2} + \sigma_{\mathrm{e}}^{2} \big)  \sum_{i=1}^{K}q_{i} + 1$, where
$\sigma_{\mathrm{e}}^{2} = N_{\mathrm{t}}^{-1}\sigma_{\mathrm{e},k}^{2}$.
Since the actual channel state $\mathbf{H}$ does not depend on the SNR, we have
$\|\mathbf{h}_{k} \|^{2},\|\widehat{\mathbf{h}}_{k} \|^{2}  = O(1)$.
It follows that $\bar{T}_{k}  \leq O\big(P_{\mathrm{t}}^{\max\{a_{k},\bar{a}_{k}\}}\big)$, from which we write
\begin{equation}
\label{Eq_DoF_part_1}
\mathrm{E}_{\mathrm{H} \mid \widehat{\mathrm{H}}} \big\{ \log_{2}(T_{k}) \mid \widehat{\mathbf{H}}  \big\}  \leq {\max\{a_{k},\bar{a}_{k}\}} \log_{2}(P_{\mathrm{t}}) + O(1).
\end{equation}
From the isotropic property and \cite[Lemma 1]{Yang2013}, we write
$\mathrm{E}_{\mathrm{H} \mid \widehat{\mathrm{H}}} \big\{  \log_{2}(I_{k}) \mid \widehat{\mathbf{H}} \big\} \geq \log_{2}(2^{\kappa} \sigma_{\mathrm{e}}^{2} \lambda_{1} + 1) + O(1)$,
where $\kappa \triangleq \mathrm{E}_{\mathrm{H} \mid \widehat{\mathrm{H}}}
\left\{ \log_{2} \left( \frac{|\mathbf{e}_{1}^{T}\widetilde{\mathbf{h}}_{k}|^{2}}{\sigma_{\mathrm{e}}^{2}} \right) \mid \widehat{\mathbf{H}}  \right\}$ is bounded, and
$\lambda_{1}$ is the dominant eigenvalue of $\sum_{i \neq k} \mathbf{p}_{i}\mathbf{p}_{i}^{H}$.
Since the maximum is lower-bounded by the average, we have
$\lambda_{1} \geq N_{\mathrm{t}}^{-1} \sum_{i \neq k} q_{i} = O\big(P_{\mathrm{t}}^{\bar{a}_{k}}\big)$ from which we obtain
\begin{equation}
\label{Eq_DoF_part_2}
\mathrm{E}_{\mathrm{H} \mid \widehat{\mathrm{H}}}  \big\{ \log_{2}(I_{k}) \mid \widehat{\mathbf{H}} \big\} \geq (\bar{a}_{k} - \alpha)^{+}\log_{2}(P_{\mathrm{t}}) + O(1).
\end{equation}
Form \eqref{Eq_DoF}, \eqref{Eq_AR_k_DoF_proof}, \eqref{Eq_DoF_part_1} and \eqref{Eq_DoF_part_2}, we have $\bar{d}_{k} \leq \max\{a_{k},\bar{a}_{k}\} - \max\{\bar{a}_{k} - \alpha,0\} $.
When $a_{k}\geq \bar{a}_{k}$, the upper-bound is given by $\min\{\alpha + a_{k} -  \bar{a}_{k},a_{k}\} $, otherwise we have
$0$.
Combining the two cases, we write $\bar{d}_{k} \leq \min \left\{ (\alpha + a_{k} -  \bar{a}_{k})^{+},a_{k} \right\} $, from which the sum DoF is upper-bounded by
\begin{equation}
\label{Eq_NoRS_SumDoF_UB_1}
\sum_{k=1}^{K} \bar{d}_{k}  \leq  \sum_{k=1}^{K} \min \left\{ (\alpha + a_{k} -  \bar{a}_{k})^{+},a_{k} \right\}.
\end{equation}
To obtain the maximum upper-bound for \eqref{Eq_NoRS_SumDoF_UB_1}, we define $\mathcal{J} \subseteq \mathcal{K}$ as the subset composed of all users with non-zero DoF.
Due to the symmetry in the CSIT qualities, it is sufficient to assume, without loss of generality, that $\mathcal{J} \triangleq \{1,\ldots, J\}$ with $J \triangleq |\mathcal{J}|$.
We define $\bar{d}(J) \triangleq \sum_{j =1}^{J} \bar{d}_{j}$.
It is evident that $\bar{d}(1) \leq 1$.
On the other hand, we have $\bar{d}(J) \leq J\alpha$ for $J > 1$. This is shown from
\begin{subequations}
\label{Eq_NoRS_SumDoF_J_UB}
\begin{align}
\label{Eq_NoRS_SumDoF_J_UB_1}
\sum_{k=1}^{J} \bar{d}_{k}  & \leq   J \alpha + \sum_{k=1}^{J} (a_{k} -  \bar{a}_{k}) \\
\label{Eq_NoRS_SumDoF_J_UB_2}
                            & \leq  J \alpha + (a_{J} -  a_{1}) + \sum_{k=1}^{J-1} (a_{k} -  a_{k+1})
\end{align}
\end{subequations}
where \eqref{Eq_NoRS_SumDoF_J_UB_1} follows from $\min\{x,y\} \leq x,y$, and the strict positivity of the DoF in $\mathcal{J}$. \eqref{Eq_NoRS_SumDoF_J_UB_2} follows from $\bar{a}_{k} \geq a_{j}$, $\forall j \neq k$, and is equal to $J \alpha$.
We conclude that the maximum upper-bound is obtained when $J=K$ for $\alpha \geq K^{-1}$, and $J=1$ otherwise. Hence, $\sum_{k=1}^{K} \bar{d}_{k}  \leq  \max\{1,K\alpha\}$.
This is also an upper-bound for
$ \lim_{P_{\mathrm{t}}\rightarrow\infty} \frac{\mathrm{E}_{\widehat{\mathrm{H}}} \big\{ \mathcal{R}(P_{\mathrm{t}}) \big\} }{\log_{2}(P_{\mathrm{t}})}$, which follows from the bounded convergence theorem.

This part of the proof is completed by showing that the upper-bound can be achieved by a feasible precoding scheme. In particular, a DoF of $1$ is achieved by Single User (SU) transmission in a Time Division Multiple Access (TDMA) fashion. On the other hand, $K\alpha$ is achieved using ZF-BF vectors designed using $\widehat{\mathbf{H}}$, and allocating the powers such that $q_{k} = P_{\mathrm{t}}^{\alpha}/K,$ for all $k \in \mathcal{K}$ as shown in \cite{Yang2013}.
\subsubsection{Proof of \eqref{Eq_DoF_RS}}
From the definitions of the ARs, we write
\begin{equation}
\nonumber
    \bar{R}_{\mathrm{c}} + \bar{R}_{1} \leq \bar{R}_{\mathrm{c},1} + \bar{R}_{1} =
    \mathrm{E}_{\mathrm{H} \mid \widehat{\mathrm{H}}}  \big\{ \log_{2}(T_{\mathrm{c},1}) - \log_{2}(I_{1}) \mid \widehat{\mathbf{H}} \big\}.
\end{equation}
Following the same approach used in the previous part,
we obtain
$\mathrm{E}_{\mathrm{H} \mid \widehat{\mathrm{H}}} \big\{\log_{2}(T_{\mathrm{c},1}) \mid \widehat{\mathbf{H}} \big\} \leq \log_{2}(P_{\mathrm{t}}) + O(1)$, and
$ \mathrm{E}_{\mathrm{H} \mid \widehat{\mathrm{H}}} \big\{ \log_{2}(I_{1}) \mid \widehat{\mathbf{H}} \big\} \geq (\bar{a}_{1} - \alpha)^{+}\log_{2}(P_{\mathrm{t}}) + O(1)$, from which we obtain the following upper-bound
\begin{equation}
\label{Eq_dc1_d1_UB}
  \bar{d}_{\mathrm{c}} + \bar{d}_{1} \leq  \bar{d}_{\mathrm{c},1} + \bar{d}_{1}  \leq  \min\{1 + \alpha - \bar{a}_{1},1\}.
\end{equation}
Next, we define $\bar{d}_{\mathrm{RS}} (J)  \triangleq \bar{d}_{\mathrm{c}} + \sum_{k=1}^{J} \bar{d}_{k}$ for $J$ users with strictly positive DoFs.
From \eqref{Eq_dc1_d1_UB}, we have  $\bar{d}_{\mathrm{RS}} (1) \leq 1$.
For $J = 2$, we have $\bar{d}_{\mathrm{RS}} (2) \leq 1 + \alpha - \bar{a}_{1} + a_{2} \leq 1+ \alpha$,
obtained from \eqref{Eq_dc1_d1_UB}, $\bar{d}_{2} \leq a_{2}$, and $\bar{a}_{1} \geq a_{2}$.
For $J > 2$, we have $\bar{d}_{\mathrm{RS}} (J) \leq 1 + (J-1)\alpha$ obtained by combining \eqref{Eq_dc1_d1_UB} and \eqref{Eq_NoRS_SumDoF_J_UB}.
Hence, the maximum upper-bound is $1+(K-1)\alpha$ obtained when $J=K$ regardless of $\alpha$.
It follows that this is also an upper-bound for
$ \lim_{P_{\mathrm{t}}\rightarrow\infty} \frac{\mathrm{E}_{\widehat{\mathrm{H}}} \big\{ \mathcal{R}_{\mathrm{RS}}(P_{\mathrm{t}}) \big\} }{\log_{2}(P_{\mathrm{t}})}$.
Finally, this is achieved using the DoF-motivated design in \cite{Yang2013}. The private precoders are given as shown in the previous part, while
$q_{\mathrm{c}} = P_{\mathrm{t}} - P_{\mathrm{t}}^{\alpha}$ and $\mathbf{p}_{\mathrm{c}} = \sqrt{q_{\mathrm{c}}}\mathbf{e}_{1}$.
\section{Generating Channel States}
\label{Appendix_Channel}
\setcounter{subsubsection}{0}
For a given channel realization, a normalized channel estimate denoted by $\widehat{\mathbf{H}}_{\mathrm{n}}$  is generating with entries drawn from $\mathcal{CN}(0,1)$.
Moreover, a set of $M$ normalized estimation errors defined as $\widetilde{\mathbb{H}}^{(M)}_{\mathrm{n}} \triangleq \big\{\widetilde{\mathbf{H}}^{(m)}_{\mathrm{n}} \mid m \in \mathcal{M} \big\} $ is generated in the same manner.
This is followed by constructing  $\mathbb{H}^{(M)}$, where the $m$th conditional realization is calculated as
$\mathbf{H}^{(m)}  =  \sqrt{1 - \sigma^{2}_{\mathrm{e}} } \widehat{\mathbf{H}}_{\mathrm{n}} +
\sqrt{\sigma^{2}_{\mathrm{e}} }\widetilde{\mathbf{H}}_{\mathrm{n}}^{(m)}$.
It is evident that the fading assumptions made at the beginning of Section \ref{Section_Numerical_Results} are preserved.
For $\alpha > 0$, i.e. $\sigma^{2}_{\mathrm{e}}$ relies on SNR, the set $\mathbb{H}^{(M)}$ varies with SNR.
To produce the ESR curves, averaging is carried out over $100$ realizations of $\widehat{\mathbf{H}}_{\mathrm{n}}$, where $\widetilde{\mathbb{H}}^{(M)}_{\mathrm{n}}$ is reused across realizations.
Reusing the same normalized channels across SNRs in this manner produces the smooth ESR curves shown in the figures.
\ifCLASSOPTIONcaptionsoff
  \newpage
\fi
\bibliographystyle{IEEEtran}
\bibliography{IEEEabrv,References}
\balance
\end{document}